\documentclass[a4paper, fleqn, usenatbib]{mnras}
\usepackage[fleqn]{amsmath}
\usepackage{mathtools}
\usepackage{graphicx}
\usepackage{upgreek}
\usepackage{wasysym}
\usepackage{amssymb}
\usepackage{multirow}
\usepackage{hyperref}
\usepackage[T1]{fontenc}
\usepackage{ae,aecompl}

\newcommand{\Rom}[1]{\uppercase\expandafter{\romannumeral #1}}
\newcommand{\rom}[1]{\lowercase\expandafter{\romannumeral #1}}
\newcommand{\Msun}{\mathrm{M_{\astrosun}}}
\newcommand{\To}{\text{--}}

\title[DRAGONS--galaxy sizes]{\mbox{Dark-ages reionization and galaxy formation simulation--\Rom{7}.}
The sizes of high-redshift galaxies}
\author[Liu et al.]{Chuanwu Liu$^1$\thanks{chuanwul@student.unimelb.edu.au},
Simon J. Mutch$^1$\thanks{smutch@unimelb.edu.au}, Gregory B. Poole$^1$, P. W. Angel$^1$, Alan R. Duffy$^2$,\newauthor
Paul M. Geil$^1$, Andrei Mesinger$^3$ and J. Stuart B. Wyithe$^1$\thanks{swyithe@unimelb.edu.au}\\
$^1$School of Physics, University of Melbourne, Parkville, VIC 3010, Australia\\
$^2$Centre for Astrophysics and Supercomputing, Swinburne University of Technology, PO Box 218, Hawthorn, VIC 3122, Australia\\
$^3$Scuola Normale Superiore, Piazza dei Cavalieri 7, I-56126 Pisa, Italy}

\begin{document}

\date{\today}

\pagerange{\pageref{firstpage}--\pageref{lastpage}} \pubyear{2014}

\maketitle

\label{firstpage}

\begin{abstract}
We investigate high-redshift galaxy sizes using a semi-analytic model constructed for the Dark-ages Reionization And
Galaxy-formation Observables from Numerical Simulation project. Our fiducial model, including strong feedback from supernovae and
photoionization background, accurately reproduces the evolution of the stellar mass function and UV luminosity function.
Using this model, we study the size--luminosity relation of galaxies and find that the effective radius scales with UV luminosity
as $R_\mathrm{e}\propto L^{0.25}$ at $z{\sim}5\To9$. We show that recently discovered very luminous galaxies at $z{\sim}7$
\citep{2016arXiv160505325B} and $z{\sim}11$ \citep{2016ApJ...819..129O} lie on our predicted size--luminosity relations.
We find that a significant fraction of galaxies at $z>8$ will not be resolved by \emph{JWST}, but \emph{GMT}
will have the ability to resolve all galaxies in haloes above the atomic cooling limit.
We show that our fiducial model successfully reproduces the redshift evolution of average galaxy sizes at $z>5$.
We also explore galaxy sizes in models without supernova feedback. The no-supernova feedback models produce galaxy sizes that are
smaller than observations. We therefore confirm that supernova feedback plays an important role in
determining the size--luminosity relation of galaxies and its redshift evolution during reionization.
\end{abstract}


\section{Introduction}
The evolution of galaxy size during the Epoch of Reionization (EoR) provides an additional probe for understanding galaxy
formation in the early Universe. In the hierarchical structure formation scenario \citep{1978MNRAS.183..341W}, dark matter haloes
form first, then baryonic gas cools and falls into their potential wells of to form galaxies. Within this scheme,
\citet{1980MNRAS.193..189F} studied the formation of galaxy discs. In this model, the spin of a rotationally supported galaxy disc
originates from the conservation of angular momentum during the collapse of cooling gas. Further analytic modelling by
\citet{1998MNRAS.295..319M} provided a relation between the disc scale length of a galaxy, $R_\mathrm{d}$, and the virial radius of
its dark matter halo, $R_\mathrm{vir}$ for infinitesimally thin discs with exponential surface density profiles. The disc size can be written as
\begin{equation}
R_\mathrm{d} = \frac\lambda{\sqrt{2}}\left(\frac{j_\mathrm{d}}{m_\mathrm{d}}\right) R_\mathrm{vir},
\label{eqn:rd}
\end{equation}
where $m_\mathrm{d}$ and $j_\mathrm{d}$ are the fraction of mass and angular momentum in the disc relative to the
halo and $\lambda$ is the spin parameter of the halo, which is a dimensionless measure of the angular momentum of the system.

The virial radius of a dark matter halo scales with redshift and virial velocity, $V_\mathrm{vir}$, or virial
mass, $M_\mathrm{vir}$, as
\begin{equation}
R_\mathrm{vir} = \left(\frac{G M_\mathrm{vir}}{100 H^2(z)}\right)^{1/3} = \frac{V_\mathrm{vir}}{10H(z)},
\end{equation}
where $H(z)$ is the Hubble parameter, and $H(z){\propto}(1+z)^{3/2}$ at high redshifts \citep{1992ARA&A..30..499C}.
Therefore, from Equation \ref{eqn:rd}, the proportionality of $R_\mathrm{d}$ with $R_\mathrm{vir}$ predicts
that the sizes of discs scale with redshift as $(1 + z)^{-3/2}$ at fixed circular velocity, or $(1+z)^{-1}$ at fixed
halo mass.

\begin{table}
 \caption{Observed evolution of galaxy sizes, $R_\mathrm{e}\propto (1+z)^m$ from literature, where $L^*_{z{=}3}$ corresponds
          to UV magnitude $M_\mathrm{UV}=-21.0$.}
 \label{tab:re-z}
 \centering
 \begin{tabular}{lccl}
  \hline
  \parbox{8mm}{$z$} & \parbox{12mm}{\centering $m$} & \parbox{32mm}{\centering{Sources}}\\
  \hline
  \multicolumn{3}{l}{$L=(0.3\To1)L^*_{z{=}3}$}\\
       $2\To6$ & $1.05\pm 0.21$ & \citet{2004ApJ...611L...1B}\\
       $2\To8$ & $1.12\pm 0.17$ & \citet{2010ApJ...709L..21O}\\
       $2\To12$ & $1.30 \pm 0.13$ & \citet{2013ApJ...777..155O}\\
       $2.5\To12$ & $1.24\pm 0.10$ & \citet{2015ApJ...804..103K}\\
       $0.5\To10$ & $1.10\pm 0.06$ & \citet{2015ApJS..219...15S}\\
       $5\To10$ & $1.32\pm 0.43$ & \citet{2015ApJ...808....6H}\\
  \\
  \multicolumn{3}{l}{$L=(0.12\To0.3)L^*_{z{=}3}$}\\
       $2\To8$ & $1.32\pm 0.52$ & \citet{2010ApJ...709L..21O}\\
       $2\To12$ & $1.30 \pm 0.13$ & \citet{2013ApJ...777..155O}\\
       $0.5\To10$ & $1.22\pm 0.05$ & \citet{2015ApJS..219...15S}\\
       $5\To10$ & $0.76\pm0.12$ & \citet{2015ApJ...808....6H}\\
  \hline
 \end{tabular}
\end{table}

Observations of Lyman break galaxies (LBGs) show that galaxies are more compact at higher redshift, and that average sizes
evolve with redshift as $(1+z)^{-m}$ with $m{\sim}1\To 1.5$ \citep[e.g.][]{2004ApJ...600L.107F, 2004ApJ...611L...1B, 2010ApJ...709L..21O, 2012A&A...547A..51G, 2013ApJ...777..155O, 2015ApJ...804..103K, 2015ApJ...808....6H, 2015ApJS..219...15S}.

Semi-analytic models have had considerable success studying the formation and evolution of galaxies in the past two decades
\citep[e.g.][]{1991ApJ...379...52W, 1993MNRAS.264..201K, 2000MNRAS.319..168C, 2006MNRAS.365...11C, 2006MNRAS.370..645B, 2011MNRAS.412.1828L, 2015arXiv150908473L}.
Galaxy sizes are important for semi-analytic models since the cold gas is assumed to settle in discs
where star formation occurs at a rate depending on the surface density \citep[e.g.][]{2006MNRAS.365...11C}.
Reproducing the evolution of galaxy sizes in the early and dense Universe is therefore important for semi-analytic models
of reionization. On the other hand, feedback mechanisms are already known to play an important role in suppressing star
formation in galaxies.

Using the observed size evolution and the luminosity function of galaxies, \citet{2011MNRAS.413L..38W} presented
a simple model to constrain the feedback mechanism using galaxy sizes:
\begin{equation}  
R_\mathrm{e}\propto L^{\frac{1}{3(1+a)}}(1+z)^{-m}.
\label{eqn:wyithe_and_loeb}
\end{equation}
Here $L$ is the galaxy luminosity, and $a$ and $m$ are free parameters which can be constrained using both the slope of the
galaxy luminosity function and galaxy size evolution. Feedback arising from energy release and momentum outflow could affect
the luminosity at fixed disc sizes. Based on the observed relation between size, luminosity and redshift,
\citet{2011MNRAS.413L..38W} ruled out the no-supernova feedback model with high confidence, and suggested a supernova feedback
model through the transfer of momentum. Here we improve on this analysis using a more realistic semi-analytic model.
Investigation of galaxy sizes using semi-analytic models have previously been made using galaxies in both the local and high redshift Universe
\citep[e.g.][]{2000MNRAS.319..168C, 2009MNRAS.397.1254G, 2010MNRAS.405..948S, 2015MNRAS.447..636X, 2016MNRAS.461..859S, 2016MNRAS.459.4109T}.
Our purpose-designed semi-analytic model provides a tool to study galaxy sizes during the EoR.

The semi-analytic model, \textsc{Meraxes} \citep[described in][hereafter Paper-\Rom{3}]{2016MNRAS.462..250M},
is a new purpose-built galaxy formation model designed for studying galaxy evolution during the EoR\footnote{The \textsc{Meraxes}
model is a part of the Dark-ages Reionization And Galaxy-formation Observables from Numerical Simulation (DRAGONS) project,
http://dragons.ph.unimelb.edu.au.}.
\textsc{Meraxes} includes a temporally and spatially coupled treatment of reionization, and is built upon the high resolution and
high snapshot-cadence $N$-body simulation $Tiamat$ \citep[hereafter Paper-\Rom{1}]{2016MNRAS.459.3025P}.
\textsc{Meraxes} successfully reproduces a series of high-redshift galaxy observables including the stellar mass function
(Paper-\Rom{3}) and UV luminosity function \citep[hereafter Paper-\Rom{4}]{2016MNRAS.462..235L}. In this paper, we run
simulations to investigate the size--luminosity relation, the size--stellar mass relation and the redshift evolution of galaxy
sizes at $5{<}z{<}10$. We aim to use the evolution of galaxy sizes to probe the physics of galaxy formation during the EoR.
In particular, we study how sensitive galaxy sizes are to feedback, especially from supernovae feedback during the EoR.

This paper is organized as follows.
In Section \ref{sec:model} we briefly introduce the semi-analytic model and $N$-body simulation used in this work.
In Section \ref{sec:size-luminosity} we study the relation between sizes and UV luminosities of galaxies.
In Section \ref{sec:resolving_galaxies} we discuss the probability of resolving galaxies using \emph{HST}, \emph{JWST} and \emph{GMT}.
In Section \ref{sec:size-mass} we study the size--stellar mass relation of model galaxies.
In Section \ref{sec:size_evolution} we present the redshift evolution of galaxy sizes and compare this with observations.
In Section \ref{sec:measure_of_sizes} we discuss the interpretation of our model sizes in the context of recent high-redshift observations.
In Section \ref{sec:summary}, we present our conclusions. Throughout this work, we employ a standard spatially-flat
$\Lambda$CDM cosmology based on \textit{Planck} 2015 data \citep{2015arXiv150201589P}:
$(h, \Omega_{\rm{m}}, \Omega_{\rm{b}}, \Omega_\Lambda, \sigma_8, n_{\rm{s}})=(0.678, 0.308, 0.0484, 0.692, 0.815, 0.968)$.
All magnitudes in this paper are presented in the AB system \citep{1983ApJ...266..713O}. The unit of luminosity, $L^*_{z=3}$,
is the characteristic luminosity at $z{\sim}3$, which corresponds to $M_\mathrm{1600}=-21.0$ \citep{1999ApJ...519....1S}.

\section{Simulation and Modelling}\label{sec:model}
The galaxy formation model used in this work is \textsc{Meraxes} (Paper-\Rom{3}).
\textsc{Meraxes} is implemented upon dark matter halo merger trees generated from the cosmological $N$-body simulation \emph{Tiamat}
(Paper-\Rom{1}). \emph{Tiamat} and \textsc{Meraxes} have special features designed for the study of reionization.

\subsection{$N$-body simulation}
The collisionless $N$-body simulation, \emph{Tiamat}, was run using a modified version of \textsc{Gadget-2} \citep{2005MNRAS.364.1105S}
and the \textit{Planck} 2015 cosmology \citep{2015arXiv150201589P}. It includes $2160^3$ particles in a comoving $100 \mathrm{Mpc}$
cube box. The mass of each particle is $2.64{\times}10^6h^{-1}\Msun$, which allows us to identify the low mass dark matter
haloes close to the hydrogen cooling limit across the redshifts relevant to reionization. Dark matter halo finding was carried out
using \textsc{Subfind} code \citep{2001MNRAS.328..726S}. This code first identifies dark matter collapsed regions
by a friends-of-friends (FoF) algorithm using a link length criterion of $0.2$ times of the mean inter-particle separation.
The self-bound substructures are subsequently identified within these FoF groups as locally overdense collections of
dark matter particles, removing unbound particles through an unbinding procedure. A FoF group typically contains a
central halo holding most of the virial mass and a group of lower-mass subhaloes which trace the undigested parts of
merger events.

\emph{Tiamat} outputs include $100$ snapshots from $z=35$ to $z=5$ with a temporal resolution of $11$ Myr per snapshot.
This high cadence resolves the dynamical time of galaxy discs at high redshift, and is comparable to the lifetime
of massive stars. Dark matter halo merger trees constructed from \emph{Tiamat} are stored and processed in a ``horizontal" form.
This allows the semi-analytic model to implement a self-consistent calculation of feedback from reionization on low mass galaxy
formation. This is achieved by incorporating the semi-numerical reionization algorithm \textsc{21cmFAST} \citep{2011MNRAS.411..955M}
at each snapshot.

\subsection{Semi-analytic model}
\textsc{Meraxes} is a new semi-analytic model based on \citet{2006MNRAS.365...11C} with updated physics for
application to $z{>}6$. It consists of baryonic infall, gas cooling, star formation, stellar mass recycling, metal enrichment,
galaxy mergers, gas stripping, and feedback from both supernova and reionization. To model the formation and evolution of galaxies
during the EoR, \textsc{Meraxes} incorporates several improvements in the feedback scheme. Firstly, it considers a delayed supernova
feedback mechanism. In an instantaneous feedback scheme, a massive star instantly produces a supernova and so releases energy and mass
within the same snapshot that the progenitor star formed. This is appropriate at low redshift, where the stellar lifetime is
short compared to the galaxy dynamical time. However, our $Tiamat$ merger trees have a much higher time resolution ${\sim}11$ Myr,
which is shorter than the lifetime of the least massive Type II supernova progenitor stars (e.g., ${\sim}40$ Myr for $8\,\Msun$ stars).
Therefore, \textsc{Meraxes} implements a delayed supernova feedback scheme, where a supernova may explode several snapshots
after the progenitor star formed. \textsc{Meraxes} also includes feedback from a spatially and temporarily variable ultraviolet
background (UVB). The UVB radiation heats the intergalactic medium and reduces baryonic infall within
small dark matter haloes suppressing both gas cooling and star formation. To achieve this, \textsc{Meraxes} integrates
the semi-numerical code \textsc{21cmFAST} \citep{2011MNRAS.411..955M} to construct the reionization structure.

We assume a standard \citet{1955ApJ...121..161S} initial mass function (IMF) with stellar mass in the range of $0.1{<}m_*{<}120\,\Msun$:
\begin{equation}
\phi(m_*)\propto m_*^{-2.35}.
\end{equation}
The free parameters in \textsc{Meraxes} were calibrated to replicate the observed stellar mass functions at $z{\sim}5$--$7$
\citep{2011ApJ...735L..34G, 2014MNRAS.444.2960D, 2015A&A...575A..96G, 2016ApJ...825....5S} and the \textit{Planck} optical
depth to electron scatting measurements \citep{2015arXiv150201589P}. For a more detailed description of \textsc{Meraxes},
see Paper-\Rom{3}.

\subsection{Disc sizes and star formation}
In our semi-analytic model, we adopt the disc scale radius from \citet{1998MNRAS.295..319M} as shown in Equation \ref{eqn:rd},
and the standard assumption $j_\mathrm{d}/m_\mathrm{d}=1$ \citep{1980MNRAS.193..189F}, for which the specific angular
momentum of the material forming the disc is the same as that of the host halo.

The spin parameter, $\lambda$, is calculated from the $N$-body simulation using the definition
\citep{2001MNRAS.321..559B}:
\begin{equation}
\lambda = \frac{J_\mathrm{vir}}{\sqrt{2}M_\mathrm{vir}V_\mathrm{vir}R_\mathrm{vir}},
\label{eqn:lambda}
\end{equation}
where $M_\mathrm{vir}$ and $J_\mathrm{vir}$ are the mass and angular momentum enclosed within the virial radius\footnote{
$R_\mathrm{vir}$ is defined as that within which the mean density is $\Delta=18\uppi^2{+}82(\Omega_m(z)-1){-}39(\Omega_m(z){-}1)^2$
times the critical density, $\rho_c$ \citep{1998ApJ...495...80B}.}, $R_\mathrm{vir}$,
and $V_\mathrm{vir}=\sqrt{GM_\mathrm{vir}/R_\mathrm{vir}}$ is the circular velocity at $R_\mathrm{vir}$. \citep[see][for
a discussion of spin parameters for haloes in \emph{Tiamat}]{2016MNRAS.459.2106A}.

Equation \ref{eqn:rd} was obtained assuming a simple model in which dark matter haloes have singular spherical isothermal
density profiles and the gravitational effects of baryonic discs are neglected. It is therefore important to note that
inclusion of gravity from the disc may alter the size and rotation curve of galaxies and modify the
dark matter concentration in the inner region of the halo. However,
\citet{1998MNRAS.295..319M} showed that a more realistic model with NFW halo profiles \citep{1997ApJ...490..493N} and
self-gravitating discs results in only minor modifications to Equation \ref{eqn:rd}.

Simulations also show that inclusion of the self-gravity of discs will lead to
instabilities of gas and stars, which drives disc material towards the centre of galaxies and results in
instability-driven star bursts and bulge growth in galaxies. This will have an impact on the distribution of disc sizes
\citep[e.g.][]{2000MNRAS.319..168C, 2006MNRAS.370..645B, 2016MNRAS.461..859S, 2016MNRAS.459.4109T}.
Another significant assumption in the model is $j_\mathrm{d}/m_\mathrm{d}{=}1$, since lots of angular momentum in the
gas component lost during galaxy assembly would lead to a smaller disc. On the other
hand, strong feedback mechanisms which release the energy and angular momentum to the interstellar medium will suppress
the formation of small discs.

To quantify these effects in semi-analytic models, \citet{2016MNRAS.461.3457G} compared galaxy sizes from semi-analytic
models \textsc{L-galaxies} and \textsc{Galform} at $z{<}2$.
\textsc{Galform} includes the self-gravity of discs while \textsc{L-galaxies} ignores it. \citet{2016MNRAS.461.3457G} showed that
self-gravity does not significantly affect the sizes of galaxies with $M_*{<}10^{9.5}\,\Msun$. However, for galaxies with
$M_*{>}10^{9.5}\,\Msun$, self-gravity of discs in \textsc{Galform} reduces galaxy sizes and results in a decreasing size--mass relation.
In this work, which considers the small galaxies that drive reionization, we do not have a large number of galaxies with
$M_*>M^{9.5}\,\Msun$ at $z>6$. Thus, we utilize the simple model of \citet{1998MNRAS.295..319M} in this study, as has been
common in semi-analytic models \citep[e.g.][]{2006MNRAS.365...11C, 2007MNRAS.375....2D}.

From Equations \ref{eqn:rd} \& \ref{eqn:lambda}, we see that the disc sizes of galaxies are determined by the properties of
dark matter haloes. We assume star formation and feedback processes do not directly modify the disc sizes. On the other hand,
the size of the disc does play a fundamental role in the build up of stellar mass.
In our model, freshly accreted baryonic matter in dark matter haloes is initially in the form of hot gas, and is assumed to
follow a singular isothermal sphere density profile. The cold gas, which cools from the hot gas
reservoir of the host FoF group, is assumed to fall onto the galaxy hosted by the central halo. \textsc{Meraxes} assumes the cold gas
settles in a rotationally supported disc with an exponential surface density profile.
Based on the observational work of \citet{1998ApJ...498..541K}, the global star formation rate of spiral galaxies can be
related to the surface density of cold gas above a given threshold. In our model, we adopt a critical surface density for
the disc, above which gas cannot maintain stability and will start forming stars. The critical density at a radius $r$ is
adopted from \citet{1996MNRAS.281..475K},
\begin{equation}
\Sigma_\mathrm{crit}(r) = \Sigma_\mathrm{norm}\left(\frac{V_\mathrm{vir}}{\mathrm{km\, s^{-1}}}\right)\left(\frac{r}{\mathrm{kpc}}\right)^{-1}
                          \Msun\mathrm{pc}^{-2},
\end{equation}
where $\Sigma_\mathrm{norm} = 0.2$ is a free parameter in \textsc{Meraxes}. Stars are assumed to form within a maximum radius
set to $R_\mathrm{disc} = 3 R_\mathrm{d}$ based on the properties of the Milky Way \citep{2000glg..book.....V}. 
By integrating $\Sigma_\mathrm{crit}$ to $R_\mathrm{disc}=R_\mathrm{d}$, we obtain the critical mass of the disc,
\begin{equation}
m_\mathrm{crit} = 2\uppi\Sigma_\mathrm{norm}\left(\frac{V_\mathrm{vir}}{\mathrm{km\, s^{-1}}}\right)
  \left(\frac{R_\mathrm{disc}}{\mathrm{kpc}}\right) 10^6\,\Msun.
\end{equation}
If the mass of cold gas in the disc, $m_\mathrm{cold}$, exceeds this threshold mass the stars will form with a star formation rate
given by
\begin{equation}
\dot{m}_* = \alpha_\mathrm{SF}\frac{m_\mathrm{cold}-m_\mathrm{crit}}{t^\mathrm{disc}_\mathrm{dyn}},
\end{equation}
where $\alpha_\mathrm{SF}=0.03$ is a free parameter describing the star formation efficiency and
$t^\mathrm{disc}_\mathrm{dyn}=R_\mathrm{d}/V_\mathrm{vir}$ is the dynamical time of the disc.

Galaxy mergers can also trigger a strong burst of star formation. We assume a fraction of the total cold gas
of the newly formed system is consumed during such a burst \citep{2001MNRAS.320..504S}
\begin{equation}
e_\mathrm{burst} = \alpha_\mathrm{burst}\left(\frac{m_\mathrm{small}}{m_\mathrm{big}}\right)^{\gamma_\mathrm{burst}},
\end{equation}
where $m_\mathrm{small}/m_\mathrm{big}$ is the mass ratio of merging galaxies, and $\alpha_\mathrm{burst} = 0.56$ and
$\gamma_\mathrm{burst} = 0.7$ are chosen to fit the numerical results of \citet{2004ApJ...607L..87C}
and \citet{1994ApJ...431L...9M, 1996ApJ...464..641M} for merger mass ratio in the range $0.1 \To 1.0$ \citep{2006MNRAS.365...11C}.
For simplicity, we assume the merger-driven burst occurs within a
single snapshot, which is comparable to the disc dynamical time of the majority of galaxies. We do not consider
irregular morphologies during galaxy mergers and the sizes of remnants are calculated using Equation \ref{eqn:rd}.

Through the star formation process, disc size affects a number of galaxy properties, including UV luminosities.
The size--luminosity relation therefore becomes an important predictor from galaxy formation models.
We note that the star forming process is
rather complicated. It is not only determined by the galaxy sizes but also by other effects including cooling, mergers and feedback.
To study the role of supernova feedback in the build up of the size-luminosity (stellar mass) relation, we also run a simulation
with the supernova feedback switched off. This no supernova model cannot reproduce the stellar mass function in detail, but is recalibrated
to provide the observed stellar mass density at $z = 5$ (see Paper-\Rom{3}).

In this paper, to compare with observations we present the sizes of model galaxies using the physical effective radius
(i.e. half-light radius), $R_\mathrm{e}$, within which half of the galaxy's luminosity originates. Here $R_\mathrm{e}$
is estimated using $R_\mathrm{e}{=}1.678 R_\mathrm{d}$, where the constant originates from the assumed exponential surface
density profile and constant mass-to-light ratio.

\subsection{UV luminosities}
Luminosity is the most direct observable of high-redshift galaxies. We calculate the UV luminosities using stellar population
synthesis. For each galaxy we obtain the stellar population components by tracking its star formation and merger history. We
integrate the stellar populations with model spectral energy distributions (SEDs) calculated using
\textsc{Starburst99} \citep{1999ApJS..123....3L, 2005ApJ...621..695V, 2010ApJS..189..309L, 2014ApJS..212...14L}
with a constant metallicity of $Z = 0.05\mathrm{Z}_\odot$. We do not include nebular components as they would not affect the UV
luminosities of our model galaxies at these redshifts.

To obtain the observed luminosities we apply a dust extinction model to each galaxy. We adopt a luminosity dependent
dust model \citep[e.g.][]{2012ApJ...756...14S, 2015ApJ...803...34B} which is based on the IRX-$\beta$ relation
from \citet{1999ApJ...521...64M} and the observed luminosity-$\beta$ relation from \citet{2014ApJ...793..115B}.
This dust model is empirical and is calibrated to reproduce the observed properties of galaxies.
For more details about the galaxy photometric modeling see Paper-\Rom{4}.

\section{Size--luminosity relation}\label{sec:size-luminosity}
\begin{figure*}
\begin{minipage}{0.99\textwidth}
    \includegraphics[width=\columnwidth]{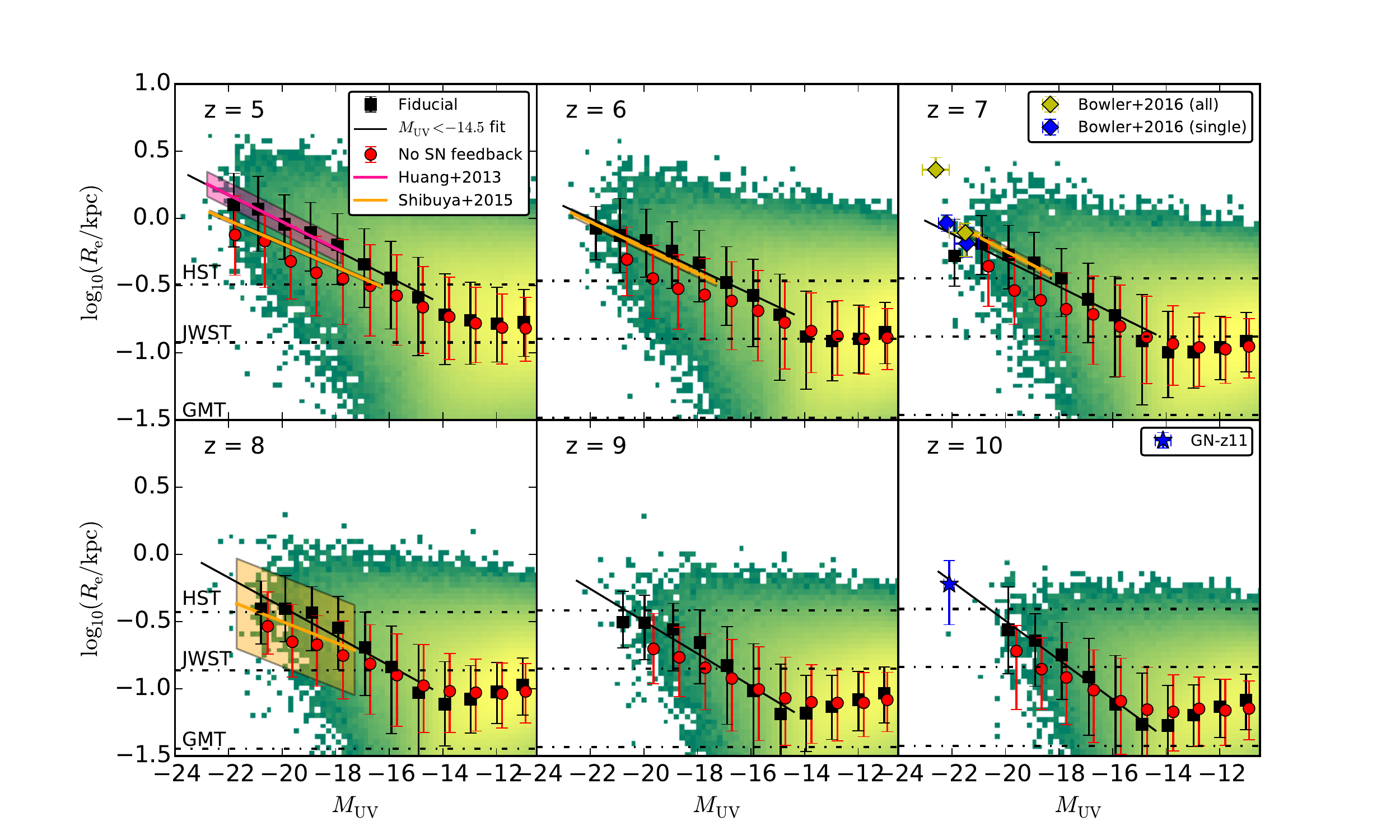}
    \caption{\label{fig:l-mg}
    Effective radius of galaxies as a function of UV luminosity at $z{\sim}5\To10$. The colour profile shows the logarithm density
    of the distribution. The black squares and error bars represent the median and $16^\mathrm{th}$ to $84^\mathrm{th}$ percentiles of
    the $R_\mathrm{e}$ distribution in bins which contain at least ten galaxies. The black solid lines are the linear best-fits
    for galaxies with $M_{1600}{<}{-14.5}$, and are extended to brighter luminosities.
    The pink and orange lines and associated shaded regions show the observed relations
    from \citet{2013ApJ...765...68H} and \citet{2015ApJS..219...15S}. The blue and yellow diamonds show the observations at $z{\sim}7$ from
    \citet{2016arXiv160505325B}. The blue star shows luminous galaxy GN-z11 found by
    \citet{2016ApJ...819..129O}. For model comparison, the red circles and error bars show the median and
    distribution of size--luminosity from the model with supernova feedback turned off. The dash-dotted lines represent
    the minimum measurable effective radii of \emph{HST}, \emph{JWST} and \emph{GMT}.}
    \label{fig:re-muv}
\end{minipage}
\end{figure*}
We first investigate the relationship between the physical size and UV luminosity of model galaxies. Fig. \ref{fig:re-muv}
shows the relation between the effective radius and UV magnitude $M_\mathrm{UV}$ for model galaxies at
$z{\sim}5\To 10$, where the UV magnitude $M_\mathrm{UV}$ is the dust-extincted luminosity at the rest-frame $1600$ \AA.
We see that at $M_\mathrm{UV}{\lesssim}{-14}$, galaxies with brighter UV luminosity tend to have larger sizes.

We note that the effective radius does not significantly change with luminosity for the galaxies with luminosities $M_\mathrm{UV}{>}{-14}$.
This is because galaxies fainter than $M_\mathrm{UV}{\sim}{-14}$ are located in the dark matter haloes of the minimum gas
cooling mass. This is similar to the turnover at $M_\mathrm{UV}{\sim}{-14}$ in the relation between UV luminosity and
the mass of dark haloes found in Paper-\Rom{4}. We see that at fixed luminosity, the size of
galaxies grows from $z{\sim}10\To 5$. We discuss the redshift evolution of galaxy sizes further in Section \ref{sec:size_evolution}.

For comparison with our simulations we show the observed $R_\mathrm{e}$--$M_\mathrm{UV}$ relations from \citep{2013ApJ...765...68H}
at $z{\sim}5$ and \citet{2015ApJS..219...15S} at $z{\sim}5\To8$, where the latter is calculated by us using the sizes data
from \citet{2015ApJS..219...15S}. Our results are in close agreement with the observations.

Recently, \citet{2016ApJ...819..129O} found an unexpectedly luminous galaxy (GN-z11) at $z{\sim}11$, which has $M_\mathrm{UV}={-22.1}{\pm}0.2$
and $R_\mathrm{e}=0.6{\pm}0.3$ kpc. In \citet{2016MNRAS.463.3556M} we demonstrated that the properties of GN-z11 are in good
agreement with the results of our model in terms of stellar mass, star formation rate and UV luminosities. We show the measured
size of GN-z11 in Fig. \ref{fig:re-muv} and we find that it is in agreement with our fitted size--luminosity relation at $z{\sim}10$.

The relation between the galaxy size and luminosity is commonly fitted by
\begin{equation}
R_\mathrm{e} = R_0\left(\frac{L_\mathrm{UV}}{L_0}\right)^\beta,
\label{eqn:re-luv}
\end{equation}
where $R_0$ is the effective radius at $L_0$, and $\beta$ is the slope. We set $L_0 = L^*_{z=3}$ which
corresponds to $M_0 = {-21}$ \citep{1999ApJ...519....1S}. This equation can be rewritten as
\begin{equation}
\log_{10} R_\mathrm{e} = -0.4\times\beta (M_\mathrm{UV} + 21) + \log_{10}(R_0).
\label{eqn:logre-muv}
\end{equation}
We linearly fit the $\log_{10}(R_\mathrm{e})\To M_\mathrm{UV}$ relation for galaxies brighter than $M_\mathrm{UV}{=}{-14.5}$
at each redshift. The best-fitting values for $R_0$ and $\beta$ at $z{\sim}5\To10$ are shown in Table \ref{tab:re-muv}.
\begin{table}
 \caption{The best-fitting parameters $R_0$ and $\beta$ (Equation \ref{eqn:re-luv}) for the
          model galaxies with UV magnitudes $M_\mathrm{UV}{<}{-14}$ at $z{\sim}5\To10$.}\label{tab:re-muv}
 \centering
 \begin{tabular}{lcc}
  \hline
  \parbox{5mm}{$z$} & \parbox{28mm}{\centering $R_0$/kpc} & \parbox{28mm}{\centering $\beta$}\\
  \hline
        $5$ & $1.17 \pm 0.05$ & $0.25 \pm 0.02$\\ 
        $6$ & $0.80 \pm 0.05$ & $0.23 \pm 0.02$\\
        $7$ & $0.61 \pm 0.07$ & $0.25 \pm 0.04$\\
        $8$ & $0.53 \pm 0.07$ & $0.28 \pm 0.04$\\
        $9$ & $0.42 \pm 0.06$ & $0.30 \pm 0.04$\\
        $10$& $0.45 \pm 0.04$ & $0.36 \pm 0.03$\\
  \hline
 \end{tabular}
\end{table}

We see that the slope of the size-luminosity relation, $\beta$, does not significantly change at $z{\sim}5\To 9$ and has a median
value of $\beta{\sim}0.25$ for galaxies with UV luminosity brighter than $M_\mathrm{UV}{\sim}{-14}$. This value agrees with observational
studies for both local and high-redshift galaxies. For example,
\citet{2000ApJ...545..781D} found $\beta = 0.253{\pm}0.020$ for local spiral galaxies. \citet{2003MNRAS.343..978S} derived
a slope of $\beta{\approx}0.26$ for the late-type galaxies from SDSS. \citet{2007ApJ...671..203C} obtained $\beta=0.321\pm0.010$ from
local field and cluster spiral galaxies. \citet{2012A&A...547A..51G} found $\beta=0.3\To 0.5$ for LBGs at $z{\sim}7$, while
\citet{2015ApJ...808....6H} derived $\beta=0.24\pm0.06$ using the \citet{2012A&A...547A..51G} data. In addition,
\citet{2013ApJ...765...68H} found $\beta=0.22$ and $0.25$ for the galaxies in GOODS and HUDF fields at $z{\sim}4$ and $z{\sim}5$
respectively. Finally \citet{2015ApJS..219...15S} investigated the galaxy effective radius from a large \emph{Hubble Space Telescope (HST)}
sample and obtained $\beta = 0.27\pm0.01$ at $z{\sim}0\To8$. They also showed that $\beta$ does not significantly evolve over
this redshift range.

Due to limitations in sample volumes and selection biases, observed values of
$\beta$ often have large uncertainties and vary between studies. For example, observations are generally biased towards
galaxies with high surface brightness and are not sensitive to measured properties of fainter, more spatially extended galaxies.
Because a model does not suffer from these selection effects and can have a large sample of both bright
and faint galaxies, we are able to investigate the true scatter of the size--luminosity relation.

The size-luminosity relation fitted to the model predictions is also consistent with the analytic prediction (Equation \ref{eqn:wyithe_and_loeb})
of \citet{2011MNRAS.413L..38W}. In that work they considered a supernova feedback model where supernova-driven winds
conserve momentum in the interaction with the galactic gas. The model results in a luminosity scaling of $a=1/3$ which
corresponds to $R_\mathrm{e}\propto L^{0.25}$. While the model without supernova feedback yields $a=0$ which corresponds
to $R_\mathrm{e}\propto L^{0.33}$.

To study the role of supernova feedback on the build up of galaxy sizes, we show the size-luminosity relation for the no supernova
feedback model in Fig. \ref{fig:re-muv} (red circles). The size-luminosity relation for the no supernova feedback model is also
flat at $M_\mathrm{UV}{>}{-14}$. This is because the minimum size is set by the mass scale of efficient cooling in both models.
There is no clear difference between the fiducial and no supernova feedback model at $M_\mathrm{UV}{>}{-17}$, where the
accumulated effect from supernova feedback on star formation histories is not significant enough to be observed. However, at
$M_\mathrm{UV}{<}{-17}$, the median size of galaxies from the no supernova feedback model is notably smaller than
the fiducial model. In other words, for the same size galaxy, the no supernova feedback model results in a much brighter luminosity.
We note that removing supernova feedback allows more stars to form, and so the model has been recalibrated to produce the correct
stellar mass density at $z=5$. The luminosity difference is ${\sim}2\To 3$ mag at $z=5\To 7$, which is larger than the
${\sim}1$ mag difference at $z = 8\To10$. This is also due to the correct galaxy mass only being achieved at $z=5$ in this
recalibrated model.

The different size-luminosity relations from these two models arise because the supernova feedback in the fiducial model
suppresses star formation resulting in a more gradual star-formation history. In contrast, galaxies without supernova feedback have much
burstier star-formation histories and contain more young stellar populations which are UV bright. These effects are more
significant at lower redshift due to the longer star-formation histories. We also ran a simulation with both supernova
and reionization feedback mechanisms switched off. However, we found the result to be almost identical to the no supernova
feedback model, with only a tiny difference at lower redshifts ($z{\sim}5\To6$).

\section{Resolving galaxies with \emph{HST}, \emph{JWST} and \emph{GMT}}\label{sec:resolving_galaxies}
The spatial resolution of a telescope with effective diameter $D_\mathrm{tel}$ is
\begin{equation}
\Delta l = \Delta \theta d_\mathrm{A} = \frac{1.22\lambda}{D_\mathrm{tel}} d_\mathrm{A},
\label{eqn:delta_l}
\end{equation}
where $\Delta \theta$ is the angular resolution determined by the Rayleigh criterion, $\lambda=1600(1+z)$ \AA\, is the observed
wavelength of UV photons and $d_\mathrm{A}$ is the angular diameter distance. In Equation \ref{eqn:delta_l}, the observed
wavelength is scaled by a factor of $(1+z)$ at fixed intrinsic wavelength, the angular diameter distance
decreases at a similar rate at $z{\gtrsim}1$. Thus the spatial resolution does not rapidly change with redshift.
Galaxy sizes are usually measured through light profile fitting \citep[e.g.][]{2002AJ....124..266P}. As a result,
one can trace the galaxy outskirt light, and obtain an effective radius bellow the spatial resolution of the telescope.
The minimum observable size of a disc depends on many galaxy properties such as the light profile and the image depth.
The comparison between the observed $R_\mathrm{e}$ and the spatial resolution limits of the \emph{Hubble Space Telescope (HST)}
indicates that values of $R_\mathrm{e}$ can be measured which are smaller than the resolution limit of the telescope by roughly
a factor of $\sim 2$ \citep[e.g.][]{2013ApJ...777..155O, 2015ApJS..219...15S}.

In Fig. \ref{fig:re-muv} we show the minimum observable disc size $R_\mathrm{min}$ of \emph{HST}, \emph{James Webb Space Telescope (JWST)},
and the \emph{Giant Magellan Telescope (GMT)}, where we adopt the relation $R_\mathrm{min}\approx \Delta l/2$ as discussed above.
We see that \emph{HST} ($D_\mathrm{tel}{=}{2.4}\,\mathrm{m}$) can resolve the $R_\mathrm{min}$ of observed galaxies at $z{\sim}5\To7$,
and the structures of typical $z{>}8$ galaxies can not be resolved. The larger diameter \emph{JWST}
($D_\mathrm{tel}{=}6.5\,\mathrm{m}$) will resolve the $R_\mathrm{min}$ for
galaxies brighter than $M_\mathrm{UV}=(-14, -16, -18)$ at $z=(6, 8, 10)$. However, with an exposure time $t_\mathrm{exp} = 10^6\,\mathrm{s}$,
\emph{JWST} will observe galaxies to $M_\mathrm{UV}=(-15.0, -15.8, -16.3)$ with signal-to-noise ratio $S/N{=}10$ at these redshifts,
hence a significant fraction of $z>8$ galaxies will be still unresolved. Due to the large mirror size, $GMT$ ($D_\mathrm{tel}{=}25\,\mathrm{m}$)
will have the ability to resolve all galaxies in haloes above the atomic cooling limit.

\section{mass--size relation}\label{sec:size-mass}
\begin{figure}
    \includegraphics[width=1.1\columnwidth]{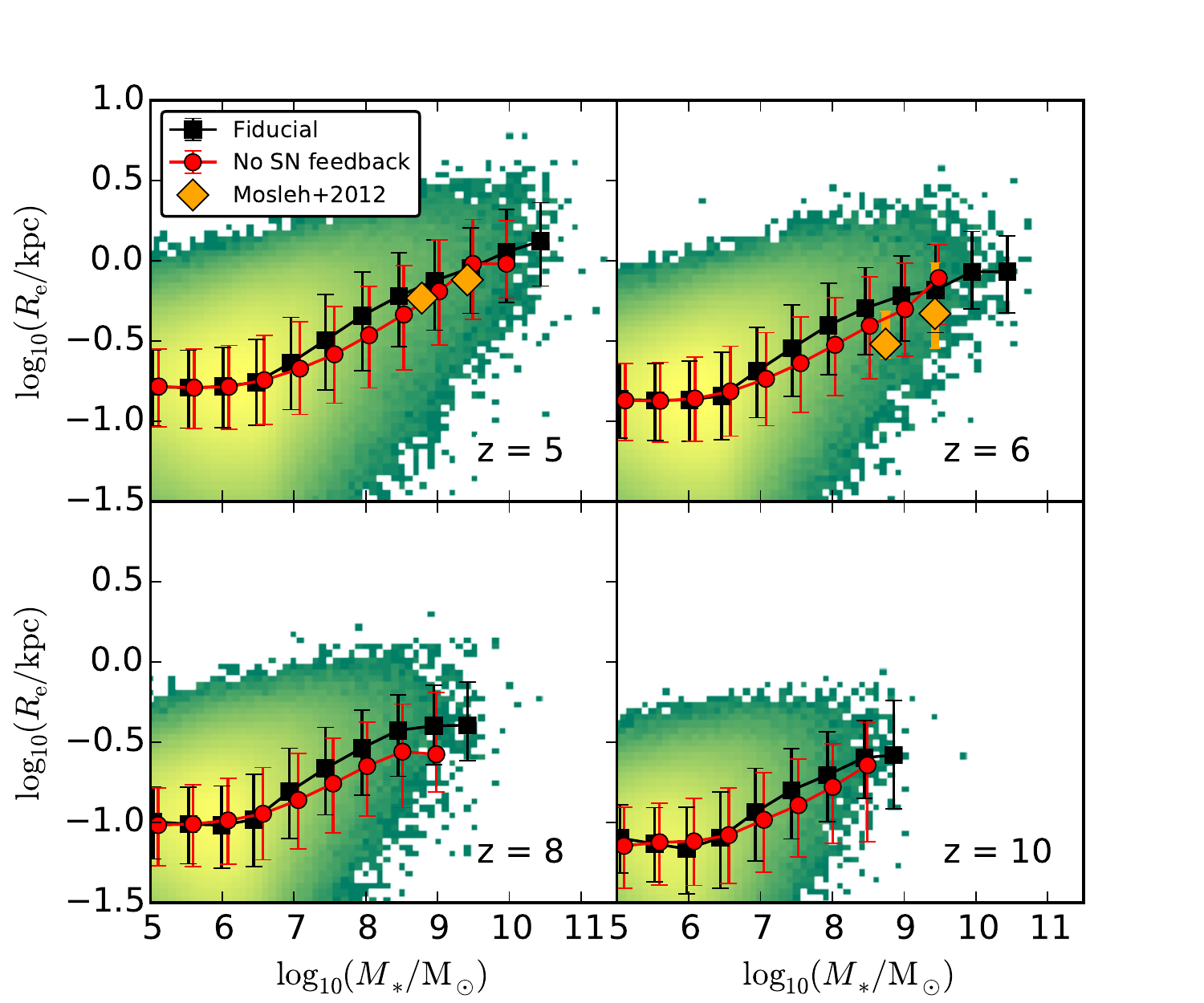}
    \caption{\label{fig:l-mstar}
    Size-mass relation of model galaxies at $z=5$, $6$, $8$, $10$. The colour profile shows the logarithm density
    of the distribution. The black squares and red circles show the median relation in bins which contain at least ten galaxies.
    The error bars represent the median and $16^\mathrm{th}$ to $84^\mathrm{th}$ percentiles of the intrinsic scatter.
    The orange diamonds show the observations from \citet{2012ApJ...756L..12M}}.
    \label{fig:re-mstar}
\end{figure}

Fig. \ref{fig:re-mstar} shows the relation between the effective radius and stellar mass of galaxies at $z{\sim}5$, $6$, $8$ and $10$
for both fiducial and no supernova feedback models. Observed data from \citet{2012ApJ...756L..12M} are also shown. The model
size-mass relation is in good agreement with these observations. We see that for galaxies with stellar masses above $10^{6.5}\,\Msun$,
more massive galaxies tend to have larger sizes. The galaxies from the fiducial model have larger sizes than the galaxies from
no supernova feedback model at fixed stellar mass. However, the difference in the size-mass relation between the fiducial and no supernova feedback
model is much smaller than in the size--luminosity relation. This is expected because we have tuned both models to produce the
galaxy stellar mass density. However, star formation histories including supernovae lead to less variable UV luminosities
resulting in larger difference seen in Fig. \ref{fig:re-muv}. For galaxies with $M_*{<}10^{6.5}\,\Msun$, our two models show similar
galaxy sizes due to the inefficient star formation in the minimum cooling mass, as was the case in the size--luminosity relation
in Fig. \ref{fig:re-muv}.

\section{Redshift evolution of sizes}\label{sec:size_evolution}

\begin{figure}
    \includegraphics[width=1.05\columnwidth]{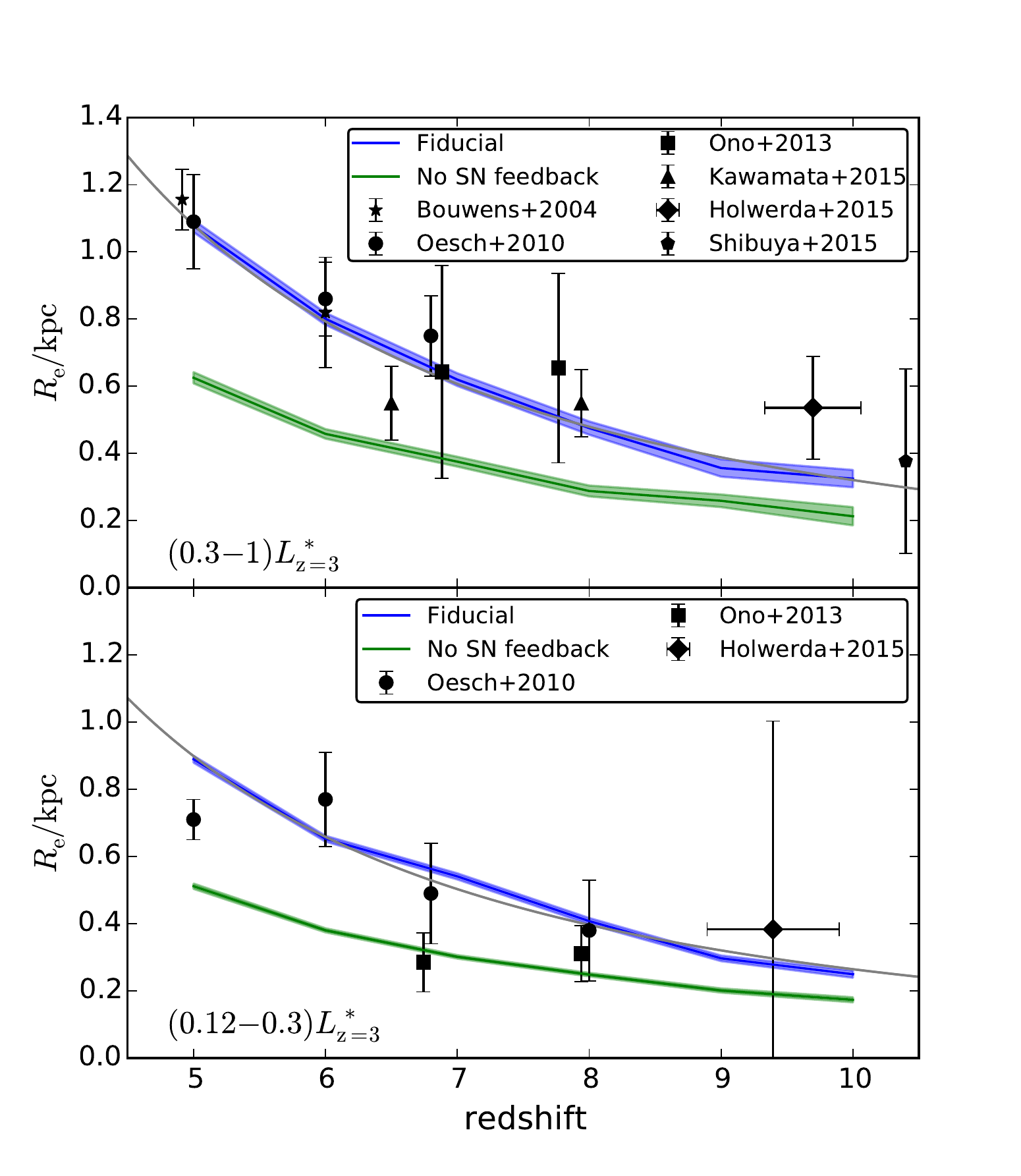}
    \caption{The redshift evolution of the mean effective radius for galaxies in the luminosity range
             $(0.3\To1)L^*_{z=3}$ (upper panel) and $(0.12\To0.3)L^*_{z=3}$ (lower panel). The blue line shows the mean
             effective radius from the fiducial model and the green line shows the mean effective radius from the model
             without supernova feedback. The shaded regions show the associated $1\sigma$ uncertainties of the means.
             The grey solid lines show the power law fit to our model.
             For comparison, we show the observed mean sizes from \citet{2004ApJ...611L...1B}, \citet{2010ApJ...709L..21O},
             \citet{2013ApJ...777..155O}, \citet{2015ApJ...804..103K}, \citet{2015ApJ...808....6H} and \citet{2015ApJS..219...15S}.
             We see that our fiducial model agrees with observations, while the no supernova model significantly underestimates the
             galaxy sizes.}
    \label{fig:re_z}
\end{figure}

The redshift evolution of galaxy sizes provides another important measurement in addition to the luminosity dependence
\citep[e.g.][]{2004ApJ...600L.107F, 2004ApJ...611L...1B, 2010ApJ...709L..21O, 2012A&A...547A..51G, 2013ApJ...777..155O, 2015ApJ...804..103K, 2015ApJ...808....6H, 2015ApJS..219...15S}.
Fig. \ref{fig:re_z} shows the redshift evolution of the effective radius predicted by our model. To compare with observations of
size evolution, galaxies were selected using their luminosity in ranges of
$(0.3\To1)L^*_{z=3}$ and $(0.12\To0.3)L^*_{z=3}$. These luminosity ranges correspond to UV magnitudes from $-21.0$ to $-19.7$
and from $-19.7$ to $-18.7$ respectively. Both fiducial and no supernova feedback models are shown in the figure.
For comparison the observed galaxy sizes from \citet{2004ApJ...611L...1B}, \citet{2010ApJ...709L..21O}, \citet{2013ApJ...777..155O},
\citet{2015ApJ...804..103K}, \citet{2015ApJ...808....6H} and \citet{2015ApJS..219...15S} are also shown.

We see that the evolution of galaxy sizes from our fiducial model is in good agreement with observations. However, the galaxy sizes in
the no supernova feedback model are underestimated at each redshift. For example, sizes at fixed luminosity in the no supernova
feedback model are ${\sim}60$ ($70$) percent of those in the fiducial model at $z{\sim}5$ ($10$). This corresponds to surface brightness
densities which are ${\sim}3$ ($2$) times larger than the fiducial model prediction. These are distinguishable differences.
To investigate the influence of parameter calibration in the no-supernova model, we have also run an
uncalibrated no-supernova feedback simulation and find a qualitatively similar result. Therefore, we conclude
that the galaxy size evolution provides an additional observable for determining the importance of supernova feedback in early
galaxy formation.

\begin{figure*}
\begin{minipage}{0.7\textwidth}
    \includegraphics[width=1.0\columnwidth]{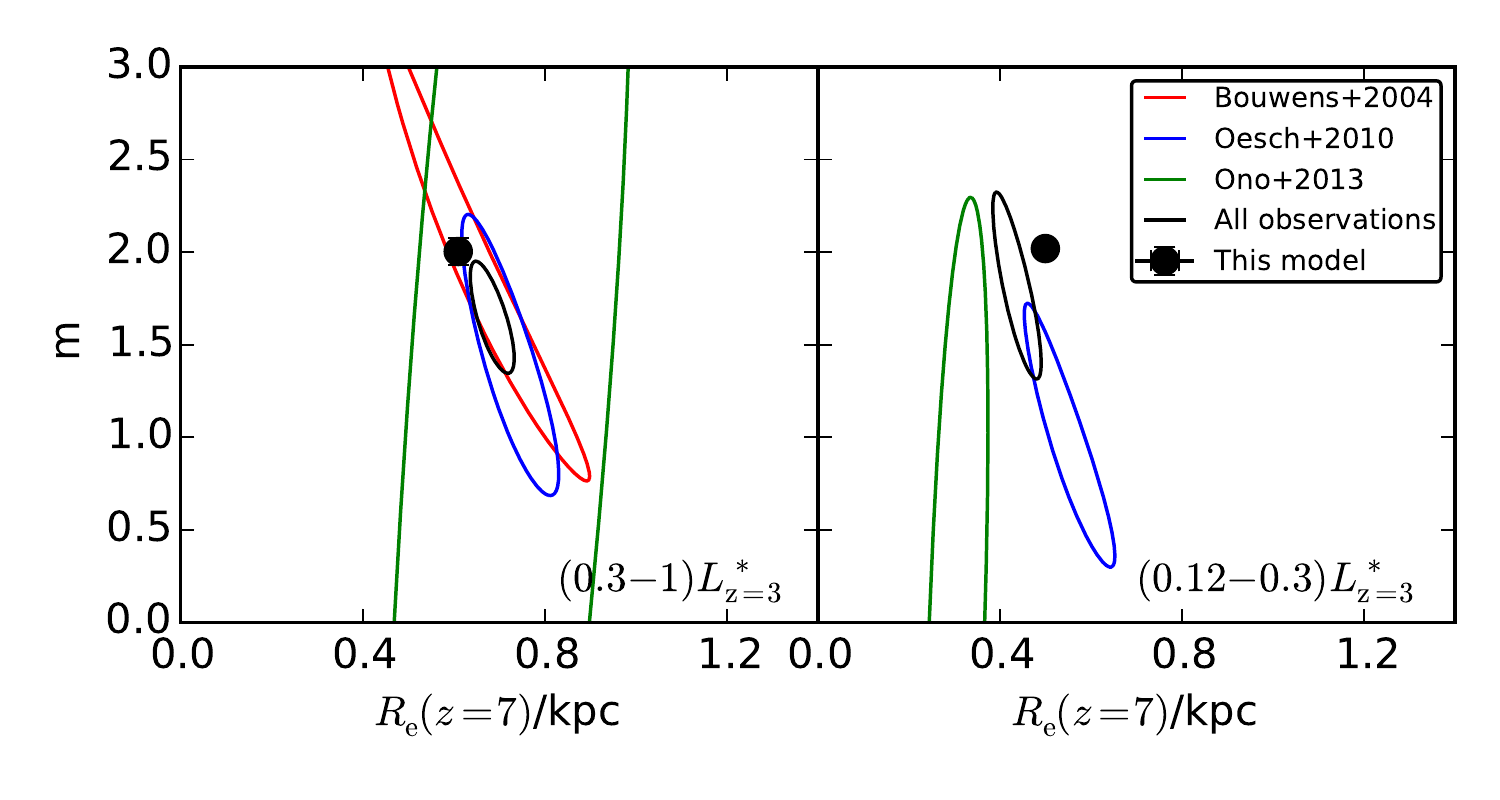}
    \caption{Confidence ellipses with $\Delta \chi^2 = 1$, which projects $1\sigma$ uncertainties on $m$ and $R_\mathrm{e}$ axes.
             The red, blue and green contours are $z{\gtrsim}5$ only observations from \citet{2004ApJ...611L...1B},
             \citet{2010ApJ...709L..21O} and \citet{2013ApJ...777..155O} respectively. The black contours are from all observations
             shown in Fig. \ref{fig:re_z}. Our best-fitting values are shown as black filled circles.}
    \label{fig:re_chi2}
\end{minipage}
\end{figure*}
We fit the model size evolution at $z{\sim}5\To 10$ using $R_\mathrm{e}\propto (1+z)^{-m}$ and find $m = 2.00{\pm}0.07$
with $R_\mathrm{e}(z{=}7)=0.61{\pm}0.01$ kpc for galaxies with luminosity in the range $(0.3\To 1)L^*_{z=3}$ and
$m = 2.02{\pm}0.04$ with $R_\mathrm{e}(z{=}7)=0.50{\pm}0.01$ kpc for galaxies with luminosity in the range $(0.12-0.3)L^*_{z=3}$.
The fitted relations are shown as grey solid lines in Fig. \ref{fig:re_z}. We also show $\Delta \chi^2 = 1$ confidence
intervals using the observations from \citet{2004ApJ...611L...1B}, \citet{2010ApJ...709L..21O} and \citet{2013ApJ...777..155O},
as well as combined observations from all data shown in Fig. \ref{fig:re_chi2}. Here we only include the observational
data at $z{>}5$ and do not include more precise measurements at $z{<}5$ which could dominate the fit.

We see that the fitted $m$ from our model is comparable to observations. For example,
$m = 1.64{\pm}0.30$ and $m = 1.82{\pm}0.51$ are derived using the combined observations shown in Fig. \ref{fig:re_chi2}
with luminosities in the ranges $(0.3\To 1)L^*_{z=3}$ and $(0.12\To0.3)L^*_{z=3}$ respectively. We note that the fits from
our model as well as $z>5$ observations give larger values for $m$ compared to observations that includes $z<5$ data as
shown in Table \ref{tab:re-z}. This may suggest that galaxy sizes undergo faster evolution at $z{>}5$ compared to the
evolution at lower redshift.

The normalization $R_\mathrm{e}(z{=}7)$ for model galaxies with luminosity in the range $(0.12\To0.3)L^*_{z=3}$ is slightly
larger than the combined observations. However, these $z>5$ observations are also inconsistent with each other due to the
large uncertainties from the small sample. We find that $R_\mathrm{e}(z{=}7)$ is in agreement with combined observations
with $3\sigma$ uncertainty.

\section{Measure of galaxy size}\label{sec:measure_of_sizes}
Before concluding, we discuss the applicability of $R_\mathrm{d}$ as a measure of galaxy size.
In observations, morphologies of LBGs are often found to be irregular and clumpy, sometimes showing multiple components
\citep[e.g.][]{1996ApJ...470..189G, 2006ApJ...652..963R, 2016MNRAS.457..440C, 2016ApJ...821...72S, 2015ApJ...800...39G, 2016arXiv160505325B}.
This could be due to two different formation mechanisms: (\rom{1}) galaxy interactions, such as mergers
\citep[e.g.][]{2006ApJ...636..592L, 2008ApJ...677...37O}; (\rom{2}) distributed and clumpy star formation regions within the same
collapsing cloud due to instabilities \citep[e.g.][]{2002ApJ...568..651G, 2007ApJ...656....1L, 2009ApJ...703..785D,
2010ApJ...709L..21O, 2013ApJ...773..153J, 2016ApJ...819L...2B}.

Morphological studies at very high redshift are more challenging. \citet{2016ApJ...821...72S} investigated the evolution
of clumpy galaxies with large \emph{HST} samples and found that the clumpy fraction increases from $z{\sim}0$ to $z{\sim}1$ but
subsequently decreases from $z{\sim}1\To3$ to $z{\sim}8$. On the other hand, high-resolution cosmological simulations show that galaxies at
$z{\gtrsim}6$ are dominated by disc morphologies \citep[e.g.][]{2011ApJ...731...54P, 2011ApJ...738L..19R, 2015ApJ...808L..17F}.
For example, using the large-volume \textsc{BlueTide} simulation,
\citet{2015ApJ...808L..17F} found that at $z=8\To 10$, up to $70$ per cent of the galaxy population more massive than
$10^{10}\,\Msun$ are disc galaxies. Detailed measurement of more compact and clumpy galaxies are limited by the angular resolution of
instruments, and the origin of observed clumpy morphologies at high-redshift is still under debate.

\citet{2016arXiv160505325B} recently published size measurements for a sample of extremely luminous galaxies at $z{\sim}7$.
\citet{2016arXiv160505325B} divided the sample into two groups (single and multi-component) according to their morphologies.
The size measurements are shown as the yellow (all galaxies) and blue (single component) diamonds in Fig. \ref{fig:re-muv}.
We see that the size-luminosity relation for the single morphology galaxies is in good agreement with our model while including
clumpy morphology galaxies leads to larger sizes. This may suggest that the multi-component galaxies are
merging systems \citep{2016arXiv160505325B}. However, we are not able to rule out the clumpy-formation scenario due to the
simplification of our semi-analytic model. Also, limited by the volume and mass resolution of our $N$-body simulation,
the bright multi-component galaxies which undergo mergers will not be resolved by our model.

\section{Conclusions}\label{sec:summary}
We have used the semi-analytic model \textsc{Meraxes} to study the dependence of galaxy size on UV luminosity, stellar mass
and redshift at $z{\sim}5 \To 10$. We also studied the effect of supernova feedback on the evolution of galaxy sizes.
We show that the rotationally supported disc model generally adopted in semi-analytic models can be used to study the
sizes of high-redshift galaxies. Our primary findings are that:

\begin{enumerate}

\item The effective radius scales with UV luminosity as \mbox{$R_\mathrm{e} \propto L^{0.25}$} for galaxies with luminosity
$M_\mathrm{UV}{\lesssim}{-14}$. Galaxies with the same disc size in the no supernova feedback model have brighter UV magnitudes
than in the fiducial model.

\item Our fiducial model with strong supernova feedback successfully reproduces the redshift evolution of average galaxy sizes
at $z>5$, which is slightly steeper than $z<5$ observations. The model with no supernova feedback produces a significantly
smaller radius at fixed luminosity than the fiducial model. 

\item The recently identified luminous galaxy GN-z11 at $z{\sim}11$ \citep{2016ApJ...819..129O} lies on our model-fitted
size-luminosity relation. The fitted relation is also in agreement with the size measurements of very luminous galaxies containing single
components and with individual components of luminous multi-component systems at $z{\sim}7$ \citep{2016arXiv160505325B}.

\item A significant fraction of \emph{$z>8$} galaxies will not be resolved by \emph{JWST}. However, \emph{GMT} will
have the ability to resolve all galaxies in haloes above the atomic cooling limit.

\end{enumerate}

We conclude that galaxy sizes provide an important additional constraint on galaxy formation physics during reionization, and
that current observations of galaxy size and evolution reinforce the importance of supernova feedback. These findings are in
agreement with results based on the stellar mass function and luminosity function.

\section*{Acknowledgements}
This research was supported by the Victorian Life Sciences Computation Initiative (VLSCI), grant ref. UOM0005, on its Peak Computing Facility hosted at the University of Melbourne, an initiative of the Victorian Government, Australia. Part of this work was performed on the gSTAR national facility at Swinburne University of Technology. gSTAR is funded by Swinburne and the Australian Governments Education Investment Fund. This research programme is funded by the Australian Research Council through the ARC Laureate Fellowship FL110100072 awarded to JSBW. AM acknowledges support from the European Research Council (ERC) under the European Unions Horizon 2020 research and innovation programme (grant agreement no. 638809 AIDA).

\bibliographystyle{mnras}
\bibliography{reference/reference}

\begin{thebibliography}{}
\makeatletter
\relax
\def\mn@urlcharsother{\let\do\@makeother \do\$\do\&\do\#\do\^\do\_\do\%\do\~}
\def\mn@doi{\begingroup\mn@urlcharsother \@ifnextchar [ {\mn@doi@}
  {\mn@doi@[]}}
\def\mn@doi@[#1]#2{\def\@tempa{#1}\ifx\@tempa\@empty \href
  {http://dx.doi.org/#2} {doi:#2}\else \href {http://dx.doi.org/#2} {#1}\fi
  \endgroup}
\def\mn@eprint#1#2{\mn@eprint@#1:#2::\@nil}
\def\mn@eprint@arXiv#1{\href {http://arxiv.org/abs/#1} {{\tt arXiv:#1}}}
\def\mn@eprint@dblp#1{\href {http://dblp.uni-trier.de/rec/bibtex/#1.xml}
  {dblp:#1}}
\def\mn@eprint@#1:#2:#3:#4\@nil{\def\@tempa {#1}\def\@tempb {#2}\def\@tempc
  {#3}\ifx \@tempc \@empty \let \@tempc \@tempb \let \@tempb \@tempa \fi \ifx
  \@tempb \@empty \def\@tempb {arXiv}\fi \@ifundefined
  {mn@eprint@\@tempb}{\@tempb:\@tempc}{\expandafter \expandafter \csname
  mn@eprint@\@tempb\endcsname \expandafter{\@tempc}}}

\bibitem[\protect\citeauthoryear{{Angel}, {Poole}, {Ludlow}, {Duffy}, {Geil},
  {Mutch}, {Mesinger}  \& {Wyithe}}{{Angel} et~al.}{2016}]{2016MNRAS.459.2106A}
{Angel} P.~W.,  {Poole} G.~B.,  {Ludlow} A.~D.,  {Duffy} A.~R.,  {Geil} P.~M.,
  {Mutch} S.~J.,  {Mesinger} A.,   {Wyithe} J.~S.~B.,  2016, \mn@doi [\mnras]
  {10.1093/mnras/stw737}, \href
  {http://adsabs.harvard.edu/abs/2016MNRAS.459.2106A} {459, 2106}

\bibitem[\protect\citeauthoryear{{Behrendt}, {Burkert}  \&
  {Schartmann}}{{Behrendt} et~al.}{2016}]{2016ApJ...819L...2B}
{Behrendt} M.,  {Burkert} A.,   {Schartmann} M.,  2016, \mn@doi [\apjl]
  {10.3847/2041-8205/819/1/L2}, \href
  {http://adsabs.harvard.edu/abs/2016ApJ...819L...2B} {819, L2}

\bibitem[\protect\citeauthoryear{{Bouwens}, {Illingworth}, {Blakeslee},
  {Broadhurst}  \& {Franx}}{{Bouwens} et~al.}{2004}]{2004ApJ...611L...1B}
{Bouwens} R.~J.,  {Illingworth} G.~D.,  {Blakeslee} J.~P.,  {Broadhurst} T.~J.,
    {Franx} M.,  2004, \mn@doi [\apjl] {10.1086/423786}, \href
  {http://adsabs.harvard.edu/abs/2004ApJ...611L...1B} {611, L1}

\bibitem[\protect\citeauthoryear{{Bouwens} et~al.,}{{Bouwens}
  et~al.}{2014}]{2014ApJ...793..115B}
{Bouwens} R.~J.,  et~al., 2014, \mn@doi [\apj] {10.1088/0004-637X/793/2/115},
  \href {http://adsabs.harvard.edu/abs/2014ApJ...793..115B} {793, 115}

\bibitem[\protect\citeauthoryear{{Bouwens} et~al.,}{{Bouwens}
  et~al.}{2015}]{2015ApJ...803...34B}
{Bouwens} R.~J.,  et~al., 2015, \mn@doi [\apj] {10.1088/0004-637X/803/1/34},
  \href {http://adsabs.harvard.edu/abs/2015ApJ...803...34B} {803, 34}

\bibitem[\protect\citeauthoryear{{Bower}, {Benson}, {Malbon}, {Helly}, {Frenk},
  {Baugh}, {Cole}  \& {Lacey}}{{Bower} et~al.}{2006}]{2006MNRAS.370..645B}
{Bower} R.~G.,  {Benson} A.~J.,  {Malbon} R.,  {Helly} J.~C.,  {Frenk} C.~S.,
  {Baugh} C.~M.,  {Cole} S.,   {Lacey} C.~G.,  2006, \mn@doi [\mnras]
  {10.1111/j.1365-2966.2006.10519.x}, \href
  {http://adsabs.harvard.edu/abs/2006MNRAS.370..645B} {370, 645}

\bibitem[\protect\citeauthoryear{{Bowler}, {Dunlop}, {McLure}  \&
  {McLeod}}{{Bowler} et~al.}{2016}]{2016arXiv160505325B}
{Bowler} R.~A.~A.,  {Dunlop} J.~S.,  {McLure} R.~J.,   {McLeod} D.~J.,  2016,
  preprint, \href {http://adsabs.harvard.edu/abs/2016arXiv160505325B} {}
  (\mn@eprint {arXiv} {1605.05325})

\bibitem[\protect\citeauthoryear{{Bryan} \& {Norman}}{{Bryan} \&
  {Norman}}{1998}]{1998ApJ...495...80B}
{Bryan} G.~L.,  {Norman} M.~L.,  1998, \mn@doi [\apj] {10.1086/305262}, \href
  {http://adsabs.harvard.edu/abs/1998ApJ...495...80B} {495, 80}

\bibitem[\protect\citeauthoryear{{Bullock}, {Kolatt}, {Sigad}, {Somerville},
  {Kravtsov}, {Klypin}, {Primack}  \& {Dekel}}{{Bullock}
  et~al.}{2001}]{2001MNRAS.321..559B}
{Bullock} J.~S.,  {Kolatt} T.~S.,  {Sigad} Y.,  {Somerville} R.~S.,  {Kravtsov}
  A.~V.,  {Klypin} A.~A.,  {Primack} J.~R.,   {Dekel} A.,  2001, \mn@doi
  [\mnras] {10.1046/j.1365-8711.2001.04068.x}, \href
  {http://adsabs.harvard.edu/abs/2001MNRAS.321..559B} {321, 559}

\bibitem[\protect\citeauthoryear{{Carroll}, {Press}  \& {Turner}}{{Carroll}
  et~al.}{1992}]{1992ARA&A..30..499C}
{Carroll} S.~M.,  {Press} W.~H.,   {Turner} E.~L.,  1992, \mn@doi [\araa]
  {10.1146/annurev.aa.30.090192.002435}, \href
  {http://adsabs.harvard.edu/abs/1992ARA%26A..30..499C} {30, 499}

\bibitem[\protect\citeauthoryear{{Cole}, {Lacey}, {Baugh}  \& {Frenk}}{{Cole}
  et~al.}{2000}]{2000MNRAS.319..168C}
{Cole} S.,  {Lacey} C.~G.,  {Baugh} C.~M.,   {Frenk} C.~S.,  2000, \mn@doi
  [\mnras] {10.1046/j.1365-8711.2000.03879.x}, \href
  {http://adsabs.harvard.edu/abs/2000MNRAS.319..168C} {319, 168}

\bibitem[\protect\citeauthoryear{{Courteau}, {Dutton}, {van den Bosch},
  {MacArthur}, {Dekel}, {McIntosh}  \& {Dale}}{{Courteau}
  et~al.}{2007}]{2007ApJ...671..203C}
{Courteau} S.,  {Dutton} A.~A.,  {van den Bosch} F.~C.,  {MacArthur} L.~A.,
  {Dekel} A.,  {McIntosh} D.~H.,   {Dale} D.~A.,  2007, \mn@doi [\apj]
  {10.1086/522193}, \href {http://adsabs.harvard.edu/abs/2007ApJ...671..203C}
  {671, 203}

\bibitem[\protect\citeauthoryear{{Cox}, {Primack}, {Jonsson}  \&
  {Somerville}}{{Cox} et~al.}{2004}]{2004ApJ...607L..87C}
{Cox} T.~J.,  {Primack} J.,  {Jonsson} P.,   {Somerville} R.~S.,  2004, \mn@doi
  [\apjl] {10.1086/421905}, \href
  {http://adsabs.harvard.edu/abs/2004ApJ...607L..87C} {607, L87}

\bibitem[\protect\citeauthoryear{{Croton} et~al.,}{{Croton}
  et~al.}{2006}]{2006MNRAS.365...11C}
{Croton} D.~J.,  et~al., 2006, \mn@doi [\mnras]
  {10.1111/j.1365-2966.2005.09675.x}, \href
  {http://adsabs.harvard.edu/abs/2006MNRAS.365...11C} {365, 11}

\bibitem[\protect\citeauthoryear{{Curtis-Lake} et~al.,}{{Curtis-Lake}
  et~al.}{2016}]{2016MNRAS.457..440C}
{Curtis-Lake} E.,  et~al., 2016, \mn@doi [\mnras] {10.1093/mnras/stv3017},
  \href {http://adsabs.harvard.edu/abs/2016MNRAS.457..440C} {457, 440}

\bibitem[\protect\citeauthoryear{{De Lucia} \& {Blaizot}}{{De Lucia} \&
  {Blaizot}}{2007}]{2007MNRAS.375....2D}
{De Lucia} G.,  {Blaizot} J.,  2007, \mn@doi [\mnras]
  {10.1111/j.1365-2966.2006.11287.x}, \href
  {http://adsabs.harvard.edu/abs/2007MNRAS.375....2D} {375, 2}

\bibitem[\protect\citeauthoryear{{Dekel}, {Sari}  \& {Ceverino}}{{Dekel}
  et~al.}{2009}]{2009ApJ...703..785D}
{Dekel} A.,  {Sari} R.,   {Ceverino} D.,  2009, \mn@doi [\apj]
  {10.1088/0004-637X/703/1/785}, \href
  {http://adsabs.harvard.edu/abs/2009ApJ...703..785D} {703, 785}

\bibitem[\protect\citeauthoryear{{Duncan} et~al.,}{{Duncan}
  et~al.}{2014}]{2014MNRAS.444.2960D}
{Duncan} K.,  et~al., 2014, \mn@doi [\mnras] {10.1093/mnras/stu1622}, \href
  {http://adsabs.harvard.edu/abs/2014MNRAS.444.2960D} {444, 2960}

\bibitem[\protect\citeauthoryear{{Fall} \& {Efstathiou}}{{Fall} \&
  {Efstathiou}}{1980}]{1980MNRAS.193..189F}
{Fall} S.~M.,  {Efstathiou} G.,  1980, \mnras, \href
  {http://adsabs.harvard.edu/abs/1980MNRAS.193..189F} {193, 189}

\bibitem[\protect\citeauthoryear{{Feng}, {Di Matteo}, {Croft}, {Tenneti},
  {Bird}, {Battaglia}  \& {Wilkins}}{{Feng} et~al.}{2015}]{2015ApJ...808L..17F}
{Feng} Y.,  {Di Matteo} T.,  {Croft} R.,  {Tenneti} A.,  {Bird} S.,
  {Battaglia} N.,   {Wilkins} S.,  2015, \mn@doi [\apjl]
  {10.1088/2041-8205/808/1/L17}, \href
  {http://adsabs.harvard.edu/abs/2015ApJ...808L..17F} {808, L17}

\bibitem[\protect\citeauthoryear{{Ferguson} et~al.,}{{Ferguson}
  et~al.}{2004}]{2004ApJ...600L.107F}
{Ferguson} H.~C.,  et~al., 2004, \mn@doi [\apjl] {10.1086/378578}, \href
  {http://adsabs.harvard.edu/abs/2004ApJ...600L.107F} {600, L107}

\bibitem[\protect\citeauthoryear{{Giavalisco}, {Steidel}  \&
  {Macchetto}}{{Giavalisco} et~al.}{1996}]{1996ApJ...470..189G}
{Giavalisco} M.,  {Steidel} C.~C.,   {Macchetto} F.~D.,  1996, \mn@doi [\apj]
  {10.1086/177859}, \href {http://adsabs.harvard.edu/abs/1996ApJ...470..189G}
  {470, 189}

\bibitem[\protect\citeauthoryear{{Goldader}, {Meurer}, {Heckman}, {Seibert},
  {Sanders}, {Calzetti}  \& {Steidel}}{{Goldader}
  et~al.}{2002}]{2002ApJ...568..651G}
{Goldader} J.~D.,  {Meurer} G.,  {Heckman} T.~M.,  {Seibert} M.,  {Sanders}
  D.~B.,  {Calzetti} D.,   {Steidel} C.~C.,  2002, \mn@doi [\apj]
  {10.1086/339165}, \href {http://adsabs.harvard.edu/abs/2002ApJ...568..651G}
  {568, 651}

\bibitem[\protect\citeauthoryear{{Gonz{\'a}lez}, {Lacey}, {Baugh}, {Frenk}  \&
  {Benson}}{{Gonz{\'a}lez} et~al.}{2009}]{2009MNRAS.397.1254G}
{Gonz{\'a}lez} J.~E.,  {Lacey} C.~G.,  {Baugh} C.~M.,  {Frenk} C.~S.,
  {Benson} A.~J.,  2009, \mn@doi [\mnras] {10.1111/j.1365-2966.2009.15057.x},
  \href {http://adsabs.harvard.edu/abs/2009MNRAS.397.1254G} {397, 1254}

\bibitem[\protect\citeauthoryear{{Gonz{\'a}lez}, {Labb{\'e}}, {Bouwens},
  {Illingworth}, {Franx}  \& {Kriek}}{{Gonz{\'a}lez}
  et~al.}{2011}]{2011ApJ...735L..34G}
{Gonz{\'a}lez} V.,  {Labb{\'e}} I.,  {Bouwens} R.~J.,  {Illingworth} G.,
  {Franx} M.,   {Kriek} M.,  2011, \mn@doi [\apjl]
  {10.1088/2041-8205/735/2/L34}, \href
  {http://adsabs.harvard.edu/abs/2011ApJ...735L..34G} {735, L34}

\bibitem[\protect\citeauthoryear{{Grazian} et~al.,}{{Grazian}
  et~al.}{2012}]{2012A&A...547A..51G}
{Grazian} A.,  et~al., 2012, \mn@doi [\aap] {10.1051/0004-6361/201219669},
  \href {http://adsabs.harvard.edu/abs/2012A%26A...547A..51G} {547, A51}

\bibitem[\protect\citeauthoryear{{Grazian} et~al.,}{{Grazian}
  et~al.}{2015}]{2015A&A...575A..96G}
{Grazian} A.,  et~al., 2015, \mn@doi [\aap] {10.1051/0004-6361/201424750},
  \href {http://adsabs.harvard.edu/abs/2015A%26A...575A..96G} {575, A96}

\bibitem[\protect\citeauthoryear{{Guo} et~al.,}{{Guo}
  et~al.}{2015}]{2015ApJ...800...39G}
{Guo} Y.,  et~al., 2015, \mn@doi [\apj] {10.1088/0004-637X/800/1/39}, \href
  {http://adsabs.harvard.edu/abs/2015ApJ...800...39G} {800, 39}

\bibitem[\protect\citeauthoryear{{Guo} et~al.,}{{Guo}
  et~al.}{2016}]{2016MNRAS.461.3457G}
{Guo} Q.,  et~al., 2016, \mn@doi [\mnras] {10.1093/mnras/stw1525}, \href
  {http://adsabs.harvard.edu/abs/2016MNRAS.461.3457G} {461, 3457}

\bibitem[\protect\citeauthoryear{{Holwerda}, {Bouwens}, {Oesch}, {Smit},
  {Illingworth}  \& {Labbe}}{{Holwerda} et~al.}{2015}]{2015ApJ...808....6H}
{Holwerda} B.~W.,  {Bouwens} R.,  {Oesch} P.,  {Smit} R.,  {Illingworth} G.,
  {Labbe} I.,  2015, \mn@doi [\apj] {10.1088/0004-637X/808/1/6}, \href
  {http://adsabs.harvard.edu/abs/2015ApJ...808....6H} {808, 6}

\bibitem[\protect\citeauthoryear{{Huang}, {Ferguson}, {Ravindranath}  \&
  {Su}}{{Huang} et~al.}{2013}]{2013ApJ...765...68H}
{Huang} K.-H.,  {Ferguson} H.~C.,  {Ravindranath} S.,   {Su} J.,  2013, \mn@doi
  [\apj] {10.1088/0004-637X/765/1/68}, \href
  {http://adsabs.harvard.edu/abs/2013ApJ...765...68H} {765, 68}

\bibitem[\protect\citeauthoryear{{Jiang} et~al.,}{{Jiang}
  et~al.}{2013}]{2013ApJ...773..153J}
{Jiang} L.,  et~al., 2013, \mn@doi [\apj] {10.1088/0004-637X/773/2/153}, \href
  {http://adsabs.harvard.edu/abs/2013ApJ...773..153J} {773, 153}

\bibitem[\protect\citeauthoryear{{Kauffmann}}{{Kauffmann}}{1996}]{1996MNRAS.28%
1..475K}
{Kauffmann} G.,  1996, \mn@doi [\mnras] {10.1093/mnras/281.2.475}, \href
  {http://adsabs.harvard.edu/abs/1996MNRAS.281..475K} {281, 475}

\bibitem[\protect\citeauthoryear{{Kauffmann}, {White}  \&
  {Guiderdoni}}{{Kauffmann} et~al.}{1993}]{1993MNRAS.264..201K}
{Kauffmann} G.,  {White} S.~D.~M.,   {Guiderdoni} B.,  1993, \mnras, \href
  {http://adsabs.harvard.edu/abs/1993MNRAS.264..201K} {264, 201}

\bibitem[\protect\citeauthoryear{{Kawamata}, {Ishigaki}, {Shimasaku}, {Oguri}
  \& {Ouchi}}{{Kawamata} et~al.}{2015}]{2015ApJ...804..103K}
{Kawamata} R.,  {Ishigaki} M.,  {Shimasaku} K.,  {Oguri} M.,   {Ouchi} M.,
  2015, \mn@doi [\apj] {10.1088/0004-637X/804/2/103}, \href
  {http://adsabs.harvard.edu/abs/2015ApJ...804..103K} {804, 103}

\bibitem[\protect\citeauthoryear{{Kennicutt}}{{Kennicutt}}{1998}]{1998ApJ...49%
8..541K}
{Kennicutt} Jr. R.~C.,  1998, \mn@doi [\apj] {10.1086/305588}, \href
  {http://adsabs.harvard.edu/abs/1998ApJ...498..541K} {498, 541}

\bibitem[\protect\citeauthoryear{{Lacey}, {Baugh}, {Frenk}  \&
  {Benson}}{{Lacey} et~al.}{2011}]{2011MNRAS.412.1828L}
{Lacey} C.~G.,  {Baugh} C.~M.,  {Frenk} C.~S.,   {Benson} A.~J.,  2011, \mn@doi
  [\mnras] {10.1111/j.1365-2966.2010.18021.x}, \href
  {http://adsabs.harvard.edu/abs/2011MNRAS.412.1828L} {412, 1828}

\bibitem[\protect\citeauthoryear{{Lacey} et~al.,}{{Lacey}
  et~al.}{2015}]{2015arXiv150908473L}
{Lacey} C.~G.,  et~al., 2015, preprint, \href
  {http://adsabs.harvard.edu/abs/2015arXiv150908473L} {} (\mn@eprint {arXiv}
  {1509.08473})

\bibitem[\protect\citeauthoryear{{Law}, {Steidel}, {Erb}, {Pettini}, {Reddy},
  {Shapley}, {Adelberger}  \& {Simenc}}{{Law}
  et~al.}{2007}]{2007ApJ...656....1L}
{Law} D.~R.,  {Steidel} C.~C.,  {Erb} D.~K.,  {Pettini} M.,  {Reddy} N.~A.,
  {Shapley} A.~E.,  {Adelberger} K.~L.,   {Simenc} D.~J.,  2007, \mn@doi [\apj]
  {10.1086/510357}, \href {http://adsabs.harvard.edu/abs/2007ApJ...656....1L}
  {656, 1}

\bibitem[\protect\citeauthoryear{{Leitherer} et~al.,}{{Leitherer}
  et~al.}{1999}]{1999ApJS..123....3L}
{Leitherer} C.,  et~al., 1999, \mn@doi [\apjs] {10.1086/313233}, \href
  {http://adsabs.harvard.edu/abs/1999ApJS..123....3L} {123, 3}

\bibitem[\protect\citeauthoryear{{Leitherer}, {Ortiz Ot{\'a}lvaro}, {Bresolin},
  {Kudritzki}, {Lo Faro}, {Pauldrach}, {Pettini}  \& {Rix}}{{Leitherer}
  et~al.}{2010}]{2010ApJS..189..309L}
{Leitherer} C.,  {Ortiz Ot{\'a}lvaro} P.~A.,  {Bresolin} F.,  {Kudritzki}
  R.-P.,  {Lo Faro} B.,  {Pauldrach} A.~W.~A.,  {Pettini} M.,   {Rix} S.~A.,
  2010, \mn@doi [\apjs] {10.1088/0067-0049/189/2/309}, \href
  {http://adsabs.harvard.edu/abs/2010ApJS..189..309L} {189, 309}

\bibitem[\protect\citeauthoryear{{Leitherer}, {Ekstr{\"o}m}, {Meynet},
  {Schaerer}, {Agienko}  \& {Levesque}}{{Leitherer}
  et~al.}{2014}]{2014ApJS..212...14L}
{Leitherer} C.,  {Ekstr{\"o}m} S.,  {Meynet} G.,  {Schaerer} D.,  {Agienko}
  K.~B.,   {Levesque} E.~M.,  2014, \mn@doi [\apjs]
  {10.1088/0067-0049/212/1/14}, \href
  {http://adsabs.harvard.edu/abs/2014ApJS..212...14L} {212, 14}

\bibitem[\protect\citeauthoryear{{Liu}, {Mutch}, {Angel}, {Duffy}, {Geil},
  {Poole}, {Mesinger}  \& {Wyithe}}{{Liu} et~al.}{2016}]{2016MNRAS.462..235L}
{Liu} C.,  {Mutch} S.~J.,  {Angel} P.~W.,  {Duffy} A.~R.,  {Geil} P.~M.,
  {Poole} G.~B.,  {Mesinger} A.,   {Wyithe} J.~S.~B.,  2016, \mn@doi [\mnras]
  {10.1093/mnras/stw1015}, \href
  {http://adsabs.harvard.edu/abs/2016MNRAS.462..235L} {462, 235}

\bibitem[\protect\citeauthoryear{{Lotz}, {Madau}, {Giavalisco}, {Primack}  \&
  {Ferguson}}{{Lotz} et~al.}{2006}]{2006ApJ...636..592L}
{Lotz} J.~M.,  {Madau} P.,  {Giavalisco} M.,  {Primack} J.,   {Ferguson} H.~C.,
   2006, \mn@doi [\apj] {10.1086/497950}, \href
  {http://adsabs.harvard.edu/abs/2006ApJ...636..592L} {636, 592}

\bibitem[\protect\citeauthoryear{{Mesinger}, {Furlanetto}  \& {Cen}}{{Mesinger}
  et~al.}{2011}]{2011MNRAS.411..955M}
{Mesinger} A.,  {Furlanetto} S.,   {Cen} R.,  2011, \mn@doi [\mnras]
  {10.1111/j.1365-2966.2010.17731.x}, \href
  {http://adsabs.harvard.edu/abs/2011MNRAS.411..955M} {411, 955}

\bibitem[\protect\citeauthoryear{{Meurer}, {Heckman}  \& {Calzetti}}{{Meurer}
  et~al.}{1999}]{1999ApJ...521...64M}
{Meurer} G.~R.,  {Heckman} T.~M.,   {Calzetti} D.,  1999, \mn@doi [\apj]
  {10.1086/307523}, \href {http://adsabs.harvard.edu/abs/1999ApJ...521...64M}
  {521, 64}

\bibitem[\protect\citeauthoryear{{Mihos} \& {Hernquist}}{{Mihos} \&
  {Hernquist}}{1994}]{1994ApJ...431L...9M}
{Mihos} J.~C.,  {Hernquist} L.,  1994, \mn@doi [\apjl] {10.1086/187460}, \href
  {http://adsabs.harvard.edu/abs/1994ApJ...431L...9M} {431, L9}

\bibitem[\protect\citeauthoryear{{Mihos} \& {Hernquist}}{{Mihos} \&
  {Hernquist}}{1996}]{1996ApJ...464..641M}
{Mihos} J.~C.,  {Hernquist} L.,  1996, \mn@doi [\apj] {10.1086/177353}, \href
  {http://adsabs.harvard.edu/abs/1996ApJ...464..641M} {464, 641}

\bibitem[\protect\citeauthoryear{{Mo}, {Mao}  \& {White}}{{Mo}
  et~al.}{1998}]{1998MNRAS.295..319M}
{Mo} H.~J.,  {Mao} S.,   {White} S.~D.~M.,  1998, \mn@doi [\mnras]
  {10.1046/j.1365-8711.1998.01227.x}, \href
  {http://adsabs.harvard.edu/abs/1998MNRAS.295..319M} {295, 319}

\bibitem[\protect\citeauthoryear{{Mosleh} et~al.,}{{Mosleh}
  et~al.}{2012}]{2012ApJ...756L..12M}
{Mosleh} M.,  et~al., 2012, \mn@doi [\apjl] {10.1088/2041-8205/756/1/L12},
  \href {http://adsabs.harvard.edu/abs/2012ApJ...756L..12M} {756, L12}

\bibitem[\protect\citeauthoryear{{Mutch}, {Geil}, {Poole}, {Angel}, {Duffy},
  {Mesinger}  \& {Wyithe}}{{Mutch} et~al.}{2016a}]{2016MNRAS.462..250M}
{Mutch} S.~J.,  {Geil} P.~M.,  {Poole} G.~B.,  {Angel} P.~W.,  {Duffy} A.~R.,
  {Mesinger} A.,   {Wyithe} J.~S.~B.,  2016a, \mn@doi [\mnras]
  {10.1093/mnras/stw1506}, \href
  {http://adsabs.harvard.edu/abs/2016MNRAS.462..250M} {462, 250}

\bibitem[\protect\citeauthoryear{{Mutch} et~al.,}{{Mutch}
  et~al.}{2016b}]{2016MNRAS.463.3556M}
{Mutch} S.~J.,  et~al., 2016b, \mn@doi [\mnras] {10.1093/mnras/stw2187}, \href
  {http://adsabs.harvard.edu/abs/2016MNRAS.463.3556M} {463, 3556}

\bibitem[\protect\citeauthoryear{{Navarro}, {Frenk}  \& {White}}{{Navarro}
  et~al.}{1997}]{1997ApJ...490..493N}
{Navarro} J.~F.,  {Frenk} C.~S.,   {White} S.~D.~M.,  1997, \apj, \href
  {http://adsabs.harvard.edu/abs/1997ApJ...490..493N} {490, 493}

\bibitem[\protect\citeauthoryear{{Oesch} et~al.,}{{Oesch}
  et~al.}{2010}]{2010ApJ...709L..21O}
{Oesch} P.~A.,  et~al., 2010, \mn@doi [\apjl] {10.1088/2041-8205/709/1/L21},
  \href {http://adsabs.harvard.edu/abs/2010ApJ...709L..21O} {709, L21}

\bibitem[\protect\citeauthoryear{{Oesch} et~al.,}{{Oesch}
  et~al.}{2016}]{2016ApJ...819..129O}
{Oesch} P.~A.,  et~al., 2016, \mn@doi [\apj] {10.3847/0004-637X/819/2/129},
  \href {http://adsabs.harvard.edu/abs/2016ApJ...819..129O} {819, 129}

\bibitem[\protect\citeauthoryear{{Oke} \& {Gunn}}{{Oke} \&
  {Gunn}}{1983}]{1983ApJ...266..713O}
{Oke} J.~B.,  {Gunn} J.~E.,  1983, \mn@doi [\apj] {10.1086/160817}, \href
  {http://adsabs.harvard.edu/abs/1983ApJ...266..713O} {266, 713}

\bibitem[\protect\citeauthoryear{{Ono} et~al.,}{{Ono}
  et~al.}{2013}]{2013ApJ...777..155O}
{Ono} Y.,  et~al., 2013, \mn@doi [\apj] {10.1088/0004-637X/777/2/155}, \href
  {http://adsabs.harvard.edu/abs/2013ApJ...777..155O} {777, 155}

\bibitem[\protect\citeauthoryear{{Overzier} et~al.,}{{Overzier}
  et~al.}{2008}]{2008ApJ...677...37O}
{Overzier} R.~A.,  et~al., 2008, \mn@doi [\apj] {10.1086/529134}, \href
  {http://adsabs.harvard.edu/abs/2008ApJ...677...37O} {677, 37}

\bibitem[\protect\citeauthoryear{{Pawlik}, {Milosavljevi{\'c}}  \&
  {Bromm}}{{Pawlik} et~al.}{2011}]{2011ApJ...731...54P}
{Pawlik} A.~H.,  {Milosavljevi{\'c}} M.,   {Bromm} V.,  2011, \mn@doi [\apj]
  {10.1088/0004-637X/731/1/54}, \href
  {http://adsabs.harvard.edu/abs/2011ApJ...731...54P} {731, 54}

\bibitem[\protect\citeauthoryear{{Peng}, {Ho}, {Impey}  \& {Rix}}{{Peng}
  et~al.}{2002}]{2002AJ....124..266P}
{Peng} C.~Y.,  {Ho} L.~C.,  {Impey} C.~D.,   {Rix} H.-W.,  2002, \mn@doi [\aj]
  {10.1086/340952}, \href {http://adsabs.harvard.edu/abs/2002AJ....124..266P}
  {124, 266}

\bibitem[\protect\citeauthoryear{{Planck Collaboration} et~al.,}{{Planck
  Collaboration} et~al.}{2015}]{2015arXiv150201589P}
{Planck Collaboration} et~al., 2015, preprint, \href
  {http://adsabs.harvard.edu/abs/2015arXiv150201589P} {} (\mn@eprint {arXiv}
  {1502.01589})

\bibitem[\protect\citeauthoryear{{Poole}, {Angel}, {Mutch}, {Power}, {Duffy},
  {Geil}, {Mesinger}  \& {Wyithe}}{{Poole} et~al.}{2016}]{2016MNRAS.459.3025P}
{Poole} G.~B.,  {Angel} P.~W.,  {Mutch} S.~J.,  {Power} C.,  {Duffy} A.~R.,
  {Geil} P.~M.,  {Mesinger} A.,   {Wyithe} S.~B.,  2016, \mn@doi [\mnras]
  {10.1093/mnras/stw674}, \href
  {http://adsabs.harvard.edu/abs/2016MNRAS.459.3025P} {459, 3025}

\bibitem[\protect\citeauthoryear{{Ravindranath} et~al.,}{{Ravindranath}
  et~al.}{2006}]{2006ApJ...652..963R}
{Ravindranath} S.,  et~al., 2006, \mn@doi [\apj] {10.1086/507016}, \href
  {http://adsabs.harvard.edu/abs/2006ApJ...652..963R} {652, 963}

\bibitem[\protect\citeauthoryear{{Romano-D{\'{\i}}az}, {Choi}, {Shlosman}  \&
  {Trenti}}{{Romano-D{\'{\i}}az} et~al.}{2011}]{2011ApJ...738L..19R}
{Romano-D{\'{\i}}az} E.,  {Choi} J.-H.,  {Shlosman} I.,   {Trenti} M.,  2011,
  \mn@doi [\apjl] {10.1088/2041-8205/738/2/L19}, \href
  {http://adsabs.harvard.edu/abs/2011ApJ...738L..19R} {738, L19}

\bibitem[\protect\citeauthoryear{{Salpeter}}{{Salpeter}}{1955}]{1955ApJ...121.%
.161S}
{Salpeter} E.~E.,  1955, \mn@doi [\apj] {10.1086/145971}, \href
  {http://adsabs.harvard.edu/abs/1955ApJ...121..161S} {121, 161}

\bibitem[\protect\citeauthoryear{{Shankar}, {Marulli}, {Bernardi},
  {Boylan-Kolchin}, {Dai}  \& {Khochfar}}{{Shankar}
  et~al.}{2010}]{2010MNRAS.405..948S}
{Shankar} F.,  {Marulli} F.,  {Bernardi} M.,  {Boylan-Kolchin} M.,  {Dai} X.,
  {Khochfar} S.,  2010, \mn@doi [\mnras] {10.1111/j.1365-2966.2010.16540.x},
  \href {http://adsabs.harvard.edu/abs/2010MNRAS.405..948S} {405, 948}

\bibitem[\protect\citeauthoryear{{Shen}, {Mo}, {White}, {Blanton}, {Kauffmann},
  {Voges}, {Brinkmann}  \& {Csabai}}{{Shen} et~al.}{2003}]{2003MNRAS.343..978S}
{Shen} S.,  {Mo} H.~J.,  {White} S.~D.~M.,  {Blanton} M.~R.,  {Kauffmann} G.,
  {Voges} W.,  {Brinkmann} J.,   {Csabai} I.,  2003, \mn@doi [\mnras]
  {10.1046/j.1365-8711.2003.06740.x}, \href
  {http://adsabs.harvard.edu/abs/2003MNRAS.343..978S} {343, 978}

\bibitem[\protect\citeauthoryear{{Shibuya}, {Ouchi}  \& {Harikane}}{{Shibuya}
  et~al.}{2015}]{2015ApJS..219...15S}
{Shibuya} T.,  {Ouchi} M.,   {Harikane} Y.,  2015, \mn@doi [\apjs]
  {10.1088/0067-0049/219/2/15}, \href
  {http://adsabs.harvard.edu/abs/2015ApJS..219...15S} {219, 15}

\bibitem[\protect\citeauthoryear{{Shibuya}, {Ouchi}, {Kubo}  \&
  {Harikane}}{{Shibuya} et~al.}{2016}]{2016ApJ...821...72S}
{Shibuya} T.,  {Ouchi} M.,  {Kubo} M.,   {Harikane} Y.,  2016, \mn@doi [\apj]
  {10.3847/0004-637X/821/2/72}, \href
  {http://adsabs.harvard.edu/abs/2016ApJ...821...72S} {821, 72}

\bibitem[\protect\citeauthoryear{{Smit}, {Bouwens}, {Franx}, {Illingworth},
  {Labb{\'e}}, {Oesch}  \& {van Dokkum}}{{Smit}
  et~al.}{2012}]{2012ApJ...756...14S}
{Smit} R.,  {Bouwens} R.~J.,  {Franx} M.,  {Illingworth} G.~D.,  {Labb{\'e}}
  I.,  {Oesch} P.~A.,   {van Dokkum} P.~G.,  2012, \mn@doi [\apj]
  {10.1088/0004-637X/756/1/14}, \href
  {http://adsabs.harvard.edu/abs/2012ApJ...756...14S} {756, 14}

\bibitem[\protect\citeauthoryear{{Somerville}, {Primack}  \&
  {Faber}}{{Somerville} et~al.}{2001}]{2001MNRAS.320..504S}
{Somerville} R.~S.,  {Primack} J.~R.,   {Faber} S.~M.,  2001, \mn@doi [\mnras]
  {10.1046/j.1365-8711.2001.03975.x}, \href
  {http://adsabs.harvard.edu/abs/2001MNRAS.320..504S} {320, 504}

\bibitem[\protect\citeauthoryear{{Song} et~al.,}{{Song}
  et~al.}{2016}]{2016ApJ...825....5S}
{Song} M.,  et~al., 2016, \mn@doi [\apj] {10.3847/0004-637X/825/1/5}, \href
  {http://adsabs.harvard.edu/abs/2016ApJ...825....5S} {825, 5}

\bibitem[\protect\citeauthoryear{{Springel}}{{Springel}}{2005}]{2005MNRAS.364.%
1105S}
{Springel} V.,  2005, \mn@doi [\mnras] {10.1111/j.1365-2966.2005.09655.x},
  \href {http://adsabs.harvard.edu/abs/2005MNRAS.364.1105S} {364, 1105}

\bibitem[\protect\citeauthoryear{{Springel}, {White}, {Tormen}  \&
  {Kauffmann}}{{Springel} et~al.}{2001}]{2001MNRAS.328..726S}
{Springel} V.,  {White} S.~D.~M.,  {Tormen} G.,   {Kauffmann} G.,  2001,
  \mn@doi [\mnras] {10.1046/j.1365-8711.2001.04912.x}, \href
  {http://adsabs.harvard.edu/abs/2001MNRAS.328..726S} {328, 726}

\bibitem[\protect\citeauthoryear{{Steidel}, {Adelberger}, {Giavalisco},
  {Dickinson}  \& {Pettini}}{{Steidel} et~al.}{1999}]{1999ApJ...519....1S}
{Steidel} C.~C.,  {Adelberger} K.~L.,  {Giavalisco} M.,  {Dickinson} M.,
  {Pettini} M.,  1999, \mn@doi [\apj] {10.1086/307363}, \href
  {http://adsabs.harvard.edu/abs/1999ApJ...519....1S} {519, 1}

\bibitem[\protect\citeauthoryear{{Stevens}, {Croton}  \& {Mutch}}{{Stevens}
  et~al.}{2016}]{2016MNRAS.461..859S}
{Stevens} A.~R.~H.,  {Croton} D.~J.,   {Mutch} S.~J.,  2016, \mn@doi [\mnras]
  {10.1093/mnras/stw1332}, \href
  {http://adsabs.harvard.edu/abs/2016MNRAS.461..859S} {461, 859}

\bibitem[\protect\citeauthoryear{{Tonini}, {Mutch}, {Croton}  \&
  {Wyithe}}{{Tonini} et~al.}{2016}]{2016MNRAS.459.4109T}
{Tonini} C.,  {Mutch} S.~J.,  {Croton} D.~J.,   {Wyithe} J.~S.~B.,  2016,
  \mn@doi [\mnras] {10.1093/mnras/stw956}, \href
  {http://adsabs.harvard.edu/abs/2016MNRAS.459.4109T} {459, 4109}

\bibitem[\protect\citeauthoryear{{V{\'a}zquez} \& {Leitherer}}{{V{\'a}zquez} \&
  {Leitherer}}{2005}]{2005ApJ...621..695V}
{V{\'a}zquez} G.~A.,  {Leitherer} C.,  2005, \mn@doi [\apj] {10.1086/427866},
  \href {http://adsabs.harvard.edu/abs/2005ApJ...621..695V} {621, 695}

\bibitem[\protect\citeauthoryear{{White} \& {Frenk}}{{White} \&
  {Frenk}}{1991}]{1991ApJ...379...52W}
{White} S.~D.~M.,  {Frenk} C.~S.,  1991, \mn@doi [\apj] {10.1086/170483}, \href
  {http://adsabs.harvard.edu/abs/1991ApJ...379...52W} {379, 52}

\bibitem[\protect\citeauthoryear{{White} \& {Rees}}{{White} \&
  {Rees}}{1978}]{1978MNRAS.183..341W}
{White} S.~D.~M.,  {Rees} M.~J.,  1978, \mnras, \href
  {http://adsabs.harvard.edu/abs/1978MNRAS.183..341W} {183, 341}

\bibitem[\protect\citeauthoryear{{Wyithe} \& {Loeb}}{{Wyithe} \&
  {Loeb}}{2011}]{2011MNRAS.413L..38W}
{Wyithe} J.~S.~B.,  {Loeb} A.,  2011, \mn@doi [\mnras]
  {10.1111/j.1745-3933.2011.01027.x}, \href
  {http://adsabs.harvard.edu/abs/2011MNRAS.413L..38W} {413, L38}

\bibitem[\protect\citeauthoryear{{Xie}, {Guo}, {Cooper}, {Frenk}, {Li}  \&
  {Gao}}{{Xie} et~al.}{2015}]{2015MNRAS.447..636X}
{Xie} L.,  {Guo} Q.,  {Cooper} A.~P.,  {Frenk} C.~S.,  {Li} R.,   {Gao} L.,
  2015, \mn@doi [\mnras] {10.1093/mnras/stu2487}, \href
  {http://adsabs.harvard.edu/abs/2015MNRAS.447..636X} {447, 636}

\bibitem[\protect\citeauthoryear{{de Jong} \& {Lacey}}{{de Jong} \&
  {Lacey}}{2000}]{2000ApJ...545..781D}
{de Jong} R.~S.,  {Lacey} C.,  2000, \mn@doi [\apj] {10.1086/317840}, \href
  {http://adsabs.harvard.edu/abs/2000ApJ...545..781D} {545, 781}

\bibitem[\protect\citeauthoryear{{van den Bergh}}{{van den
  Bergh}}{2000}]{2000glg..book.....V}
{van den Bergh} S.,  2000, {The Galaxies of the Local Group}.
Cambridge

\makeatother
\end{thebibliography}

\appendix

\bsp
\label{lastpage}
\end{document}